\documentclass[useAMS,usenatbib,pdfsync,lfeqn]{mn2e}

\usepackage{amstext}
\usepackage{amsmath, float}
\usepackage{graphicx}

\newcommand       \be           {\begin{equation}}
\newcommand       \ee           {\end{equation}}
\newcommand       \bea          {\begin{eqnarray}}
\newcommand       \eea          {\end{eqnarray}}

\newcommand       \aad           {\xi}

\def\simlt{\mathrel{\hbox{\rlap{\hbox{\lower4pt\hbox{$\sim$}}}\hbox{$<$}}}}
\def\simgt{\mathrel{\hbox{\rlap{\hbox{\lower4pt\hbox{$\sim$}}}\hbox{$>$}}}}
\def\simlt{\mathrel{\hbox{\rlap{\hbox{\lower4pt\hbox{$\sim$}}}\hbox{$<$}}}}
\def\simgt{\mathrel{\hbox{\rlap{\hbox{\lower4pt\hbox{$\sim$}}}\hbox{$>$}}}}
\def\simlt{\mathrel{\hbox{\rlap{\hbox{\lower4pt\hbox{$\sim$}}}\hbox{$<$}}}}
\def\simgt{\mathrel{\hbox{\rlap{\hbox{\lower4pt\hbox{$\sim$}}}\hbox{$>$}}}}
\def\lesssim{\mathrel{\hbox{\rlap{\hbox{\lower4pt\hbox{$\sim$}}}\hbox{$<$}}}}
\def\gtrsim{\mathrel{\hbox{\rlap{\hbox{\lower4pt\hbox{$\sim$}}}\hbox{$>$}}}}

\def\simlt{\mathrel{\hbox{\rlap{\hbox{\lower4pt\hbox{$\sim$}}}\hbox{$<$}}}}
\def\simgt{\mathrel{\hbox{\rlap{\hbox{\lower4pt\hbox{$\sim$}}}\hbox{$>$}}}}

\title[A dynamo closure for 3+1 resistive GR-MHD]
{A fully covariant mean-field dynamo closure for numerical 3+1 resistive GRMHD}
\author[N. Bucciantini \& L. Del Zanna]{N. Bucciantini$^{1,3}$
\thanks{E-mail:niccolo@arcetri.inaf.it} \&  L. Del Zanna$^{2,3}$ \\
$^{1}$ INAF - Osservatorio Astrofisico di Arcetri, L.go E.~Fermi 5, 50125, Firenze, Italy\\ 
$^{2}$ Dipartimento di Fisica ed Astronomia, Universit\`a di Firenze, L.go E.~Fermi 2, 
50125, Firenze, Italy \\
$^{3}$ INFN - Sezione di Firenze, Via G.~Sansone 1, 50019 Sesto Fiorentino, Firenze, Italy
}
\begin{document}

\date{Accepted . Received ; in original form }

\pagerange{\pageref{firstpage}--\pageref{lastpage}} \pubyear{????}

\maketitle

\label{firstpage}

\begin{abstract}

The powerful high-energy phenomena typically encountered in astrophysics
invariably involve physical engines, like neutron stars and black hole accretion disks,
characterized by a combination of highly magnetized plasmas, strong gravitational
fields, and relativistic motions.
In recent years numerical schemes for General Relativistic MHD (GRMHD)
have been developed to model the multidimensional dynamics of such systems, 
including the possibility of an evolving spacetime. Such schemes have been also extended
beyond the ideal limit including the effects of resistivity, in an attempt to model 
dissipative physical processes acting on small scales (sub-grid effects) 
over the global dynamics.
Along the same lines, the magnetic field could be amplified by the presence of
turbulent dynamo processes, as often invoked to explain the high values of
magnetization required in accretion disks and neutron stars.
Here we present, for the first time, a further extension to include the possibility of 
a mean-field dynamo action within the framework of numerical $3+1$ (resistive) GRMHD. 
A fully covariant dynamo closure is proposed, in analogy with
the classical theory, assuming a simple $\alpha$-effect in the
comoving frame. Its implementation into a finite-difference scheme
for GRMHD in dynamical spacetimes [the X-ECHO code: \cite{Bucciantini_Del-Zanna11a}] 
is described, and a set of numerical test is presented and compared with
analytical solutions wherever possible. 

\end{abstract}


\section{Introduction}
\label{sec:intro}

A strong magnetic field plays a crucial role in many high-energy astrophysical
systems. It is believed to be the key element in the context of Gamma
Ray Bursts (GRBs)
\citep{Duncan_Thompson92a,Thompson94a,Meszaros_Rees97a,
Lee_Wijers+00a,Lyutikov_Blackman01a,Vlahakis_Konigl01a,
van-Putten_Levinson03a, Lyutikov06a,Komissarov_Barkov07a,
Metzger_Thompson+07a,Uzdensky_MacFadyen07a,Uzdensky_MacFadyen07b,
Barkov_Komissarov08a,
Bucciantini_Quataert+08b,Bucciantini_Quataert+09a,Lyons_OBrien+10a,
Metzger_Giannios+11a,Rezzolla_Giacomazzo+11a},
AGN-jets \citep{Blandford_Znajek77a,Blandford_Payne82a,Belvedere_Molteni84a,
Khanna_Camenzind92a,Konigl_Kartje94a,Balbus_Hawley98a,Tomimatsu00a,
Fendt_Memola01a, Sauty_Tsinganos+02a,van-Putten_Levinson03a,Duttan_Biermann07a,
Hawley08a,Penna_McKinney+10a,Tchekhovskoy_Narayan+11a}, magnetars
\citep{Duncan_Thompson92a,Thompson_Duncan96a,Murakami99a,Lyutikov03b,
Lyutikov06b,Braithwaite_Spruit06a,Woosley10a,Kasen_Bildsten10a,Yu11a},
and its dissipation and reconnection to be at the origin of many
typical high energy phenomena \citep{Uzdensky11a}. A large-scale ordered
magnetic field is in fact crucial to power many high energy systems. 
A strong magnetic field close to the central black hole (BH) in accretion disks 
is invoked to explain the launching of relativistic jets. 
The pulsar and magnetar paradigms require a strong dipolar magnetic 
field in neutron stars (NSs). Magnetic field's stresses provide an efficient
way to convert rotational or accretion energy into bulk flow, and to power relativistic winds. 

However, the origin of this strong and ordered magnetic field remains
poorly understood. In particular, the environment in which such a
magnetic field is supposed to arise is often characterized by
turbulence \citep{Zhang_MacFadyen+09a,Mizuno_Pohl+11a} 
and instabilities, ranging from MRI in accretion disks
\citep{Balbus_Hawley98a} to kink and Tayler and convection in 
NSs \citep{Miralles_Pons+00a,Miralles_Pons+02a,
Braithwaite_Spruit06a,Ruediger_Schultz+08a}. 
While turbulent small-scale motions can easily amplify a
seed field up to equipartition with the turbulent kinetic energy, one
expects such a field to be highly tangled. If a large-scale ordered
mean field can arise due to small scale velocity fluctuations is
highly debated, and understanding what its configuration and its
geometry are, and under what conditions it is realized, is of
fundamental importance.
It is not unreasonable that, during the formation of compact
objects like BHs or NSs, any frozen-in large-scale field might be
amplified due to advection. However, as an explanation of the required
magnetic field, this simply shifts the problem from the object to the
progenitor. Moreover, in BHs, NSs, and GRBs, a large amount of
angular momentum is required in the engine, but any strong and
pre-existing magnetic field would rapidly slow down rotation 
\citep{Duez_Liu+06a,Obergaulinger_Aloy+06a,
Bisnovatyi-Kogan_Moiseenko+08a,Benson_Babul09a}. 

Processes that lead to in-situ amplification of large-scale magnetic field, 
due to the dynamics of the flow, are usually referred as \emph{dynamos} 
\citep{Parker55a,Parker87a,
Moffatt78a,Burbidge_Layzer+81a,Zeldovich_Ruzmaikin87a,
Roberts_Soward92a,Kulsrud_Cowley+97a,Sato99a,Brandenburg_Subramanian05a}. 
There are many kind of dynamo processes but, essentially, all involve twisting of the magnetic
fieldlines by the flow and reconnection events that allow to rearrange
irreversibly the field topology. Dynamos  usually require a fully
three-dimensional flow structure. Cowling's theorem, for example, 
states that dynamos cannot work for two-dimensional flow patterns.

Of particular interest in astrophysics are dynamo processes due to
the presence of small-scale fluctuations in the flow and turbulence.
The idea that small-scale velocity and magnetic field fluctuations
might be correlated, leading to a large-scale effective electromotive
force capable of amplifying and generating  a large-scale magnetic
field, is at the base of the so-called {\it mean-field} dynamo theory
\citep{Moffatt78a,Krause_Raedler80a,Brandenburg_Dobler02a,
Brandenburg_Subramanian05a}. 
Mean-field dynamos have been applied to a large variety of astrophysical
systems, from the Sun \citep{Parker09a} to stellar magnetism 
\citep{Brandenburg_Dobler02a},
from accretion disks \citep{Khanna_Camenzind96b,Pariev_Colgate+07a} 
to proto-neutron stars \citep{Bonanno_Rezzolla+03a}, from the
origin of the galactic field \citep{Shukurov02a} to that of the cosmological
primordial field \citep{Kulsrud_Zweibel08a}, just to cite a few.

The formulation of a mean-field closure of Maxwell's equation in the
context of General Relativity (GR) has been done only for two
astrophysical cases, to our knowledge: 
\citet{Marklund_Clarkson05a} have investigated
the origin of the cosmic magnetic field during inflation, while
\citet{Khanna_Camenzind96a} and \citet{Khanna99a} have
focused on the specific case of disks in Kerr metric. Both have
performed some study in the limiting kinematic case, where the
back-reaction of the magnetic field on the plasma is neglected. 
\citet{Khanna99a} also have not considered the displacement current, 
which however might become non-negligible near the event horizon of a BH, 
or for rapidly evolving systems. They show that various new terms can arise
in GR which are not present in a flat space-time. 
There is also a debate if Cowling's anti-dynamo theorem still
holds in GR: frame dragging effects can generate currents even for
axisymmetric configurations (the gravito-magnetic term) in the absence
of turbulence \citep{Khanna_Camenzind96a}.

The possibility of a stable and reliable mean-field closure 
for turbulent dynamos is however very important in the context of 
numerical simulations. Quite often large-scale simulations are
required to investigate the dynamics of GRB engines, accretion disks
in AGN, and magnetosphere-jet coupling. The development of small-scale
fluctuations is a property of turbulence that make its numerical
investigation within global models prohibitive if not impossible at
the moment. The idea behind a mean-field approach is that the effects
of physical processes at scales that cannot be resolved can instead
be modeled by an appropriate closure of the equations, so that
the problem can become treatable by numerical investigation. However
we want to stress here that, in the mean-field approach, determining
what could be realistic values for the parameters
that are used in the closure, is usually non trivial, and often
requires the use of mesoscale informations, and extrapolation of flow
properties to small unresolved scales. It rests to be proved that a mean
field approach can achieve full resolution of microscopic physics on 
 the macro-scale. 

Here we present the first \emph{fully covariant} mean-field dynamo closure 
of Maxwell's equations, by extending the covariant Ohm's law
widely used in resistive GRMHD \citep{Lichnerowicz67a, Anile89a}.
The new contribution is due to an $\alpha$-dynamo term proportional
to the large-scale magnetic field as measured in the local comoving frame, 
in analogy to the classical case.
Moreover, having in mind numerical applications, the equations are then
cast in the $3+1$ formalism, leading to a modified evolution equation
for the spatial electric field as measured by Eulerian observers.
When dynamo effects are negligible, the equation reduces to that
already derived for resistive MHD in \emph{special} relativity 
\citep{Komissarov07a,Palenzuela_Lehner+09a,Dumbser_Zanotti09a},
thus extending it to the more general case of $3+1$ resistive GRMHD.
It is not our intention to discuss here either the validity of mean-field dynamo
or its theoretical implication within GRMHD.
As already done for the purely resistive case, in the above cited works,
here we mainly focus on the numerical implementation of the proposed
closure. Our numerical references are the ECHO and X-ECHO codes
\citep{Del-Zanna_Zanotti+07a,Bucciantini_Del-Zanna11a}, 
and the actual implementation and validation tests of this new 
dynamo closure will refer to these schemes.

The plan of the paper is the following.
In Sect.~\ref{sec:dyn} we discuss the mean-field closure for the
induction equation, and present our covariant GR extension.
Sect~\ref{sec:adm} is devoted to its representation in the $3+1$ formalism,
necessary for numerical modeling,
whereas in Sect.~\ref{sec:num} we discuss its actual implementation within
the X-ECHO code for GRMHD.  In Sect.~\ref{sec:res}
we present a set of simple standard tests done both in the so called {\it kinematic} and in
the fully dynamic regime, and compare them with previously published
results and with analytic and semi-analytic solutions.  Finally we present our
conclusion in Sect.~\ref{sec:con}.

In the following we assume a signature ${-, +, +, +}$ for the
space-time 
metric and we use Greek letters $\mu, \nu, \lambda, ...$  (running from 0
to 3) for 4D space-time tensor components, while Latin letters
$i, j, k, ...$  (running from 1 to 3) will be employed for 3D spatial
tensor components, and spatial vectors will be often written using bold
face characters. Moreover, we set $c = G = M_\odot = 1$ and
all $\sqrt{4\pi}$ factors will be absorbed in the definition of the electromagnetic fields.

\section{The covariant Ohm's law and its mean-field dynamo closure}
\label{sec:dyn}

Let us now briefly discuss the main 
idea behind the mean-field dynamo theory \citep{Krause_Raedler80a}.
Ohm's law for resistive (classical) MHD reads
\begin{align}
\bmath{E} + \bmath{v}\times\bmath{B} = \eta\,\bmath{J}; 
\quad \bmath{J}= \bmath{\nabla}\times \bmath{B},
\label{eq:res_mhd}
\end{align}
where, $\bmath{E}$, $\bmath{B}$, $\bmath{v}$, $\bmath{J}$ are the
electric field, the magnetic field, the velocity and the current
respectively, and $\eta$ is the resistivity or coefficient of (isotropic) magnetic diffusivity. 
If now one separates the quantities 
into their large-scale mean values $\bar{\bmath{E}}$, $\bar{\bmath{v}}$, $\bar{\bmath{B}}$, and 
small-scale fluctuating parts  $\delta\bmath{E}, \delta\bmath{v}, \delta\bmath{B}$, 
a new electromotive force due to turbulent motion appears in Ohm's law
\begin{align}
\bar{\bmath{E}}  + \bar{\bmath{v}} \times\bar{\bmath{B}} = 
- \overline{ \delta \bmath{v}\times \delta\bmath{B}} + \eta \bar{\bmath{J}}; 
\quad \bar{\bmath{J}} = \bmath{\nabla}\times \bar{\bmath{B}}.
\end{align}
In general the small-scale fluctuating quantities are correlated and thus their
product has a non-vanishing mean. The key assumption is that this mean
can be written as a function of the mean quantities and their derivatives, 
and in the simplest case that it is linear in the value of both the 
mean magnetic field and its curl. Namely, it is often assumed that 
\be
\overline{ \delta \bmath{v}\times \delta\bmath{B}} = \alpha \bar{\bmath{B}}
- \beta \nabla\times \bar{\bmath{B}}
\ee
where the two scalar coefficients are both proportional to the local turbulent
correlation time $\tau_c$. In particular
\be
\alpha = - \textstyle{\frac{1}{3}}\tau_c \,
\overline{\delta \bmath{v} \times (\nabla\times\delta\bmath{v})};
\quad \beta = \textstyle{\frac{1}{3}}\tau_c \,\overline{\delta \bmath{v}^2},
\ee
where the $\alpha$-term is proportional to the kinetic helicity and the $\beta$-term
is related to a random walk for fluid elements.
For a deeper insight on the properties of turbulence and a more exhaustive
derivation of the $\alpha$ and $\beta$ terms the reader is referred, for instance,
to \citet{Kulsrud05a}. Inclusion of magnetic helicity in the definition of $\alpha$
and a different mean-field dynamo closure allowing for a dynamical
definition of the electromotive force due to small-scale turbulent fluctuations
may be found in \citet{Blackman_Field02a}.

Dropping the bars and referring from now on to just large-scale averaged 
quantities,  the classical form for the dynamo closure is then
\begin{align}
\bmath{E} + \bmath{v}\times\bmath{B} = 
- \alpha \,\bmath{B} + (\beta + \eta)\,\bmath{J}.
\end{align}
Notice that in general both $\alpha$ and $\beta$
will be tensors, however we will focus here on the isotropic case
where they can be dealt with as scalars. Their values might depend on fluid
quantities like density, temperature, or magnetic field strength.
Moreover, even the specific physical problem at hand could have an influence,
as the resulting asymptotic turbulent state may be strongly affected by
the assumed initial conditions.
When these coefficients can be treated as constants, the 
induction equation for classical MHD becomes
\be
\partial_t\bmath{B} = \nabla\times (\bmath{v}\times\bmath{B}) + 
\alpha \,\nabla\times\bmath{B} + (\beta + \eta)\,\nabla^2\bmath{B},
\ee
and it is now apparent that the presence of the $\alpha$-term 
may introduce exponentially
growing modes that are known as mean-field dynamo waves, 
whereas the $\beta$-term acts as a sort of \emph{turbulent
  diffusivity} or  {\it turbulent resistivity}, which is often
dominant over the kinetic one (in the fast dynamo case). 
 The $\beta$-term can
be interpreted as due to turbulent mixing: the convection  cells mix up
magnetic field lines of different polarities on small scales, and thus
reduce the mean field, which is equal to the field averaged over
larger scales
In the kinematic regime, $\alpha$, $\beta$ (and $\eta$) are input
parameters, as well as $\bmath{v}$, and only the above equation
needs to be solved.

Let us finally summarize mean-field dynamo treatment within classical MHD 
by rewriting the classical Ohm's law above as
\begin{align}
\bmath{E}^\prime = \aad \,\bmath{B} + \eta\,\bmath{J}.
\label{eq:alphadyn}
\end{align}
that is by replacing
$\bmath{E} + \bmath{v}\times\bmath{B}  \to \bmath{E}^\prime$, the electric
field in the frame comoving with the fluid, $- \alpha\to\aad$ (to avoid conflict with 
the lapse function in the $3+1$ standard notation, to be introduced in next section), 
and $\beta + \eta \to \eta$, combining magnetic and turbulent diffusivity in a single coefficient.
In the remainder of this paragraph we will propose a fully covariant generalization of
Eq.~\ref{eq:alphadyn}, which is novel in the literature, to our knowledge.

The covariant Maxwell's equations are written in terms of the Faraday 
(antisymmetric) tensor $F^{\mu\nu}$ and its dual $F^{* \mu\nu} =
\frac{1}{2}\epsilon^{\mu\nu\lambda\kappa}F_{\lambda\kappa}$, where
$\epsilon^{\mu\nu\lambda\kappa}$ is the spacetime Levi-Civita pseudo-tensor
(here we use the convention $(-g)^{1/2}\epsilon^{0123}=-(-g)^{-1/2}\epsilon_{0123}=1$), as
\be
\nabla_\mu F^{*\mu\nu}=0, \quad \nabla_\mu F^{\mu\nu}=-I^{\nu},
\ee
where $I^\mu$ is the 4-current. The above quantities may be decomposed
in the reference frame comoving with the fluid 4-velocity $u^\mu$ as
\begin{align}
& F^{\mu\nu} = u^\mu e^\nu - e^\mu u^\nu + 
\epsilon^{\mu\nu\lambda\kappa}b_\lambda u_\kappa, \\
& F^{*\mu\nu} = u^\mu b^\nu - b^\mu u^\nu - 
\epsilon^{\mu\nu\lambda\kappa}e_\lambda u_\kappa,
\end{align}
and
\be
I^\mu = q_0 u^\mu + j^\mu,
\ee
where $e^\mu=F^{\mu\nu}u_\nu$, $b^\mu=F^{*\mu\nu}u_\nu$, $q_0=-I^\mu u_\mu$, 
and $j^\mu$ are, respectively, the electric field, magnetic field, charge density, 
and (conduction) current measured in such frame 
($e^\mu u_\mu = b^\mu u_\mu = j^\mu u_\mu =0$).

Ohm's law for (isotropic) resistive GRMHD is usually written as a linear relation between 
the comoving electric field and current \citep{Lichnerowicz67a, Anile89a}
\be
e^\mu = \eta\, j^\mu,
\ee
and the ideal GRMHD relation of a vanishing comoving electric field $e^\mu=0$
is recovered by letting $\eta=0$ (an ideal plasma with infinite conductivity), which
was the closure employed in \citet{Del-Zanna_Zanotti+07a} and 
\citet{Bucciantini_Del-Zanna11a}.
The straightforward extension to include a mean-field $\alpha$-dynamo effect
appears to be
\be
e^\mu=\aad\, b^\mu + \eta\, j^\mu,
\label{eq:dynrel}
\ee
where a new term proportional to the comoving mean magnetic field $b^\mu$
now appears and again only the isotropic case has been considered.
Both the coefficients $\aad$ and $\eta$  serve as a sort of sub-grid modeling
of the turbulent motions, and turbulent diffusivity is supposed to be in general
higher than its kinetic value, as in the classical case.
Eq.~\ref{eq:dynrel} is thus our proposed fully covariant generalization of
Eq.~\ref{eq:alphadyn}, to which it correctly reduces in the comoving frame
where $\bmath{v}=0$ (see next section).

We want to stress here that, as in the standard
mean-field dynamo theory, there is a large degree of freedom in the
choice of the closure relations. It can even be debated if a closure
in terms of mean quantities and their derivative can be found at all.
The quest for an appropriate closure is only apparently more complex in 
general relativity, due to the requirement of general covariance,
but practically this seems to support the validity of our simple expression
in Eq.~\ref{eq:dynrel}. However, we want to stress once again that
this can be so straightforward only when the isotropic case is assumed,
as done in this work for simplicity. Anisotropic resistive MHD in a
Minkowskian spacetime was considered by \cite{Zanotti_Dumbser11a}.
On the other hand, our scalar parameters $\xi$ and $\eta$ may be
both function of all the other (macroscopic) quantities, and thus
evolve dynamically in time with them. For simplicity, however, even
if the scheme is built to take into account this most general case, 
numerical tests will be presented only for constant $\xi$ and $\eta$.

\section{The dynamo closure in $3+1$ GRMHD}
\label{sec:adm} 

In the $3+1$ formalism the line element is usually given as
\be
ds^2 = - \alpha^2dt^2 +\gamma_{ij}(dx^i+\beta^idt)(dx^j+\beta^jdt), 
\ee
where $\alpha$ the {\it lapse} function, $\beta^i$ the spatial {\it shift vector}, 
both arbitrary due to gauge invariance in the choice of coordinates,
and $\gamma_{ij}$ is the spatial 3-metric, with determinant $\gamma$
($-g=\alpha^2\gamma)$. 
In this metric the unit vector of the Eulerian observer's 4-velocity is
\begin{align}
n_\mu = (-\alpha,0 ), \quad n^\mu = (1/\alpha,-\beta^i/\alpha),
\end{align}
and projection onto the spatial hyper-surfaces normal to $n^\mu$ 
is achieved via the 3-metric
\be
\gamma_{\mu\nu} = g_{\mu\nu} +n_\mu n_\nu.
\ee
Spatial projection for a generic vector  $V^\mu$ (or tensor) is then
$\perp V^\mu = \gamma^\mu_{\,\nu} V^\nu = V^\mu + (V^\nu n_\nu) n^\mu $,
and for a spatial vector $\perp V^\mu \equiv V^\mu$ . Such a spatial vector must 
have a vanishing contravariant temporal component, since $V^\mu n_\mu=0$.

If we now decompose the electromagnetic quantities within the $3+1$
Eulerian framework, in analogy with the previous relation we write
\begin{align}
& F^{\mu\nu} = n^\mu E^\nu - E^\mu n^\nu + 
\epsilon^{\mu\nu\lambda\kappa}B_\lambda n_\kappa, \\
& F^{*\mu\nu} = n^\mu B^\nu - B^\mu n^\nu - 
\epsilon^{\mu\nu\lambda\kappa}E_\lambda n_\kappa,
\end{align}
and
\be
I^\mu = q n^\mu + J^\mu,
\ee
where $E^\mu=F^{\mu\nu}n_\nu$, $B^\mu=F^{*\mu\nu}n_\nu$, $q$, $J^\mu$ are, respectively,
the electric field, magnetic field, charge density, and (conduction) current measured
in such frame ($E^\mu n_\mu = B^\mu n_\mu = J^\mu n_\mu =0$).
The Maxwell's equations take the usual form, plus some extra terms
due to $3+1$ GR metric
\begin{align}
& \!\gamma^{-1/2}\partial_t (\gamma^{1/2}\bmath{E}) \! - \! 
\bmath{\nabla}\times (\alpha\bmath{B} - \bmath{\beta}\times\bmath{E} ) \! = \!
- (\alpha\bmath{J}- q\bmath{\beta}),
\label{eq:ampere} \\
& \gamma^{-1/2}\partial_t (\gamma^{1/2} \bmath{B}) \! + \!
\bmath{\nabla}\times (\alpha\bmath{E} + \bmath{\beta}\times\bmath{B}) \! = \! 0,
\label{eq:faraday} \\
& \bmath{\nabla}\cdot\bmath{E} = q,
\label{eq:gauss} \\
& \bmath{\nabla}\cdot\bmath{B}= 0,
\label{eq:divb}
\end{align}
and we do not repeat here the momentum-energy conservation equation,
the continuity equation for the mass density (an equivalent one holds for the electric
charge density $q$), and Einstein's field equations in the $3+1$ formalism.
Compared to ideal GRMHD, where the electric field is a derived quantity,
we must now also consider the evolution and constraint equations for $\bmath{E}$,
where the sources $q$ and $\bmath{J}$ appear.

Let us see now how to treat the various forms of the generalized Ohm's law.
It is first convenient to decompose the quantities related to the frame comoving
with the fluid within the $3+1$ Eulerian split of time and space, namely
\begin{align}
& u^\mu = \Gamma n^\mu + \Gamma v^\mu, \\
& e^\mu = \Gamma (\bmath{E}\cdot\bmath{v}) n^\mu + 
\Gamma (E^\mu + \epsilon^{\mu\nu\lambda}v_\nu B_\lambda), \\
& b^\mu = \Gamma (\bmath{B}\cdot\bmath{v}) n^\mu + 
\Gamma (B^\mu - \epsilon^{\mu\nu\lambda}v_\nu E_\lambda), \\
& j^\mu = (q - q_0\Gamma) n^\mu + J^\mu -q_0\Gamma v^\mu,
\end{align}
where $\Gamma = (1-v^2)^{-1/2}$ is the Lorentz factor of the fluid flow,
$q_0=\Gamma (q - \bmath{J}\cdot\bmath{v})$, and $\epsilon^{\mu\nu\lambda} = 
\epsilon^{\mu\nu\lambda\kappa}n_\kappa$ is the spatial Levi-Civita pseudo-tensor,
for which $\gamma^{1/2}\epsilon^{123}=1$.
We now have three possibilities:
\begin{enumerate}
\item 
{\bf Ideal GRMHD} ($e^\mu=0$):

The spatial projection readily provides the usual ideal MHD assumption
\be
\bmath{E} + \bmath{v}\times\bmath{B}= 0,
\label{eq:evb}
\ee
exactly as in the classical limit.

\item
{\bf Resistive GRMHD} ($e^\mu = \eta j^\mu$):

The time projection leads to $\Gamma (\bmath{E}\cdot\bmath{v}) = \eta (q-q_0\Gamma)$,
which may be used to express $q_0$ in the spatial component.
The result is
\be
\Gamma [\bmath{E} + \bmath{v}\times\bmath{B}  - (\bmath{E}\cdot\bmath{v})\bmath{v}] =
\eta (\bmath{J} - q\bmath{v}),
\ee
as already found in previous treatments within \emph{special} relativistic resistive MHD
\citep{Komissarov07a,Palenzuela_Lehner+09a,Dumbser_Zanotti09a}, thus
here we extend its validity to $3+1$ GRMHD.
When $\eta=0$ we reduce to the previous ideal MHD case, while for
$|\bmath{v}|\ll 1$, $|\bmath{E}|\sim |\bmath{v}||\bmath{B}|$ we recover the classical
limit of Eq.~\ref{eq:res_mhd}. 

\item 
{\bf Resistive GRMHD + dynamo} ($e^\mu = \aad b^\mu + \eta j^\mu$):

The time projection is now
\be
\Gamma (\bmath{E}\cdot\bmath{v}) = \eta (q-q_0\Gamma)
+ \aad \Gamma (\bmath{B}\cdot\bmath{v}),
\ee
which again may be used to express $q_0$ in the spatial component.
Now the result is
\be
\Gamma [\bmath{E} + \bmath{v}\times\bmath{B}  - (\bmath{E}\cdot\bmath{v})\bmath{v}] =
\eta (\bmath{J} - q\bmath{v})
+ \aad \Gamma [\bmath{B} - \bmath{v}\times \bmath{E}  - (\bmath{B}\cdot\bmath{v})\bmath{v}] ,
\ee
which is a novel closure to our knowledge, reducing to the case of
resistive GRMHD when $\aad =0$ and to the classical case of Eq.~\ref{eq:alphadyn}
for small velocities.
\end{enumerate}
It is interesting to note that in all cases the presence of a curved or even evolving
GR metric is not apparent in Maxwell's equations, since $\alpha$ or $\beta^i$
terms do not appear explicitly, whereas $\gamma_{ij}$ is just needed to work
out scalar and cross products between (spatial) vectors.

While in resistive schemes for classical MHD [see e.g. \citet{Landi-Londrillo+08}]
$q$ and $\bmath{E}$ do not play a role and the system is closed simply by 
taking $\bmath{J}=\bmath{\nabla}\times\bmath{B}$ in Ampere's law, the presence
of Maxwell's displacement current in the relativistic case forces one to use
Eq.~\ref{eq:ampere} to evolve the electric field in time. Only in
ideal GRMHD  is it still
possible to neglect the Maxwell equations for the electric field, 
since $\bmath{E}$ is provided from the ideal Ohm's law Eq.~\ref{eq:evb}. 
In the other cases the two equations for $\bmath{E}$ must be evolved or 
preserved in time, and we need to face the problem of dealing with the unknown
sources $q$ and $\bmath{J}$. In the proposed scheme we use Ohm's law to provide
an expression for the spatial current density $\bmath{J}$, while the constraint
on $q$ is enforced by using the Gauss theorem in Eq.~\ref{eq:gauss}.
Notice that here we do not evolve the charge density via the corresponding 
continuity equation (charge conservation law), since we choose to impose directly 
$q=\bmath{\nabla}\cdot\bmath{E}$. 

In the most general case, including mean-field dynamo effects, the result is
\begin{align}
& \gamma^{-1/2}\partial_t (\gamma^{1/2}\bmath{E}) - 
\bmath{\nabla}\times (\alpha\bmath{B} - \bmath{\beta}\times\bmath{E} )
+ (\alpha\bmath{v}- \bmath{\beta}) q = \nonumber \\
& - \alpha\Gamma [\bmath{E} + \bmath{v}\times\bmath{B}  - 
(\bmath{E}\cdot\bmath{v})\bmath{v}] /\eta  \nonumber \\
& + \xi\, \alpha\Gamma [\bmath{B} - \bmath{v}\times\bmath{E}  -
 (\bmath{B}\cdot\bmath{v})\bmath{v}] /\eta.
\label{eq:stiff}
\end{align}
When $\eta= 0$ we can neglect the time evolution term and the rest of the left hand side, 
therefore time integration of the electric field is not required; 
when also $\aad=0$ we recover the
ideal Ohm's law condition $\bmath{E} = - \bmath{v}\times\bmath{B}$, as expected.
When $\eta>0$ problems arise because Eq.~\ref{eq:stiff} is usually \emph{stiff}, especially
for low values of the resistivity, since the source terms on the right are larger than
the time marching term by a large factor $1/\eta$, and some sort of \emph{implicit}
numerical scheme must be employed for time integration. Note that in this respect 
the presence of an dynamo $\alpha-$term adds no further complexity to the numerical algorithm,
and its implementation within an existing resistive code is expected to be straightforward.
 
\section{Implementation in the ECHO code}
\label{sec:num} 

Let us discuss here how the closure relation  Eq.~\ref{eq:stiff} can be
solved in a numerical scheme for $3+1$ GRMHD
taking as a reference the ECHO \citep{Del-Zanna_Zanotti+07a} and X-ECHO
\citep{Bucciantini_Del-Zanna11a} codes developed in the ideal regime. 
Here we propose a very simple method to integrate the equation implicitly.
Considering a first order discretization of the time derivative, one can
write an expression for the electric field at the end of a timestep $\Delta t$
as a function of other quantities, both at the end (superscript $^{(1)}$)
and at the beginning (superscript $^{(0)}$) of the implicit procedure.

Introducing the spatial components instead of the vectorial notation we have
\begin{align} 
& E^{i(1)}= {\gamma^{(1)}}^{-1/2} {\gamma^{(0)}}^{1/2} Q^{i(0)}-\nonumber\\
&[\Gamma^{(1)} E^{i(1)}+\epsilon^{ijk}\tilde{v}_j^{(1)}B_k^{(1)}-
(E^{k(1)}\tilde{v}_k^{(1)})\tilde{v}^{i(1)}/\Gamma^{(1)}]/\tilde{\eta} + \nonumber\\
& \aad [\Gamma^{(1)}B^{i(1)} - \epsilon^{ijk}\tilde{v}_j^{(1)}E_k^{(1)}-
(B^{k(1)}\tilde{v}_k^{(1)})\tilde{v}^{i(1)}/\Gamma^{(1)}]/\tilde{\eta} 
\label{eq:dynim}
\end{align}
where rigorously $\tilde{\eta}=\tilde{\eta}^{(1)}$ and $\aad=\aad^{(1)}$, since
the two coefficients might be in principle
function of other variables like temperature, density or magnetic field.
The other assumptions are
\begin{align}
&\tilde{v}^i = \Gamma v^i,\;\; \tilde{v}_i =
\Gamma v_i,\;\;\;\Gamma^2=1+\tilde{v}_i \tilde{v}^i,\nonumber\\
& Q^{i}= E^{i} +\Delta t [ - (\alpha v^i-\beta^i)q +
\epsilon^{ijk} \partial_j (\alpha B_k - \epsilon_{klm}\beta^l E^m) ],\label{eq:qi}\\
& q = \gamma^{-1/2} \partial_k (\gamma^{1/2}E^k), \nonumber\\
&1/\tilde{\eta}=\Delta t \,\alpha/\eta. \nonumber
\end{align}
If we now solve for $E^i$, after some lengthy algebra we find
\begin{align} 
&E^i [\Gamma +  \tilde{\eta} +
\aad^2(\Gamma^2 - 1)/(\Gamma + \tilde{\eta})]  = \nonumber\\ 
& - \epsilon^{ijk}\tilde{v}_j B_k +
\tilde{\eta}[ Q^i+ (Q^k \tilde{v}_k)\tilde{v}^i ]/ (1+\tilde{\eta} \Gamma) \nonumber\\ 
& + \aad [\Gamma B^i- \tilde{\eta} (B^k\tilde{v}_k)\tilde{v}^i/(1+\tilde{\eta}\Gamma)] \nonumber\\
& - \aad [(\Gamma^2 - 1)B^i-(\tilde{v}^k B_k)\tilde{v}^i +
\,\tilde{\eta}\epsilon^{ijk}\tilde{v}_j Q_k]/ (\Gamma + \tilde{\eta}) \nonumber\\
&+ \aad^2[\Gamma \epsilon^{ijk}\tilde{v}_j B_k]/ (\Gamma + \tilde{\eta})   \nonumber\\
& + \aad^2 [ \tilde{\eta} \,\Gamma (Q^k\tilde{v}_k)  +
\aad (B^k\tilde{v}_k)]\tilde{v}^i/ [(1+ \tilde{\eta}\Gamma)(\Gamma + \tilde{\eta})   ],
\label{eq:a10}
\end{align}
where for simplicity we have let ${\gamma^{(1)}}^{-1/2} {\gamma^{(0)}}^{1/2} Q^{i(0)}\to Q^i$
and dropped all superscripts.
When $\aad=0$, that is in the purely resistive GRMHD case, many terms cancel out
and we are left with the simple relation
\be
E^i (\Gamma +  \tilde{\eta}) = - \epsilon^{ijk}\tilde{v}_j B_k +
\tilde{\eta}[ Q^i+ (Q^k \tilde{v}_k)\tilde{v}^i ]/ (1+\tilde{\eta} \Gamma), 
\ee
which automatically includes the limit for an ideal plasma $E^i = - \epsilon^{ijk}v_j B_k$
when $\eta=0$.
 
\subsection{Primitive variables}

Eq.~\ref{eq:a10} provides the electric field at the end of a timestep
as a function of other primitive variables like the velocity and the
magnetic field. However numerical schemes for GRMHD usually evolve
conserved variables like momentum and energy, and primitive variables
are not immediately available at the end of a timestep, but must be
derived by inverting a set of non-linear equations. This implies that
Eq.~\ref{eq:a10} must be solved simultaneously with the inversion from
conserved variables to primitive. In analogy to
\cite{Palenzuela_Lehner+09a}, the derivation of the primitive
variables and the electric field is done along the following lines:

\begin{itemize}
\item Given the conserved variables at the end of a timestep, 
a guess for the pressure $p^*$ is chosen, using the value at the previous timestep.
\item With this value $p^*$ of the pressure kept fixed, a guess for the
  $v^{i*}$ is chosen, again  using the values at the previous timestep.
\item $E^i$ components are derived according to Eq.~\ref{eq:a10}. This
  step is performed only when the solution of an implicit equation is
  required.  In general the $Q^i$ contain all of the explicit terms (see
  the following subsection on time-stepping). Eq.~\ref{eq:qi} provides
  the value of $Q^i$ in the simple case for a first order implicit
  solver.
\item The momentum equations are inverted keeping fixed the value $p^*$, by
  means of a Newton-Raphson scheme, where the Jacobian is computed numerically, 
to provide a new guess $v^{i*}$  for the $v^i$ components. A new
electric field is derived and this loop is iterated until convergence.
\item The energy equation is finally inverted by
  means of a Newton-Raphson scheme using the values
  of the velocity  $v^{i*}$  and electric field obtained in the inner cycle, to
  provide a new guess  $p^*$ for the pressure. Again the derivative 
  required for the  Newton-Raphson scheme is computed numerically,
  allowing for a general EoS.
\item The overall cycle over the pressure $p$ is repeated until convergence.
\end{itemize}

This approach has the advantage to allow the use of a general
equation of state, hence it is not limited to the ideal gas or polytropic EoS,
and in principle can be generalized to other closure relations for
Ohm's law. We opted for numerical Jacobians, as opposed to approximations
to the true analytical ones \citep{Palenzuela_Lehner+09a}, which might be extremely complex to
derive and expensive to compute.

The above implementation is straightforward in the Cowling approximation, where
the metric terms are fixed in time, but it can be easily extended also to a
dynamical space-time by using for example the XCFC approach as used in X-ECHO
\citep{Cordero-Carrion_Cerda-Duran+09a,Bucciantini_Del-Zanna11a}, 
given that the equation for the evolution of the
metric terms are decoupled from the inversion from conserved to
primitive variables. The only problem is actually the
treatment of the lapse function, appearing in the definition of $\tilde{\eta}$.
While in XCFC the conformal factor and the determinant of the
three-metric $\gamma$ can be easily computed from the conserved
quantities, the lapse $\alpha$ requires the previous knowledge of the
primitive variables, including the electric field. This implies that
one should in principle solve simultaneously for $\alpha$ and the primitive variables.
Given that in general the lapse is a slowly and smoothly varying function of time,
we prefer to use for simplicity $\alpha^{(0)}$ in $\tilde{\eta}$, and
we expect this choice to introduce only minor errors. 

\subsection{Time stepping and constrained transport}

The scope of this work is to present and verify the implementation of
a dynamo closure in a numerical scheme for $3+1$ GRMHD. Our code allows
the use of two distinct time-stepping approaches: 
a simple 1st-2nd splitting of the temporal evolution, between the implicit and
the explicit part, and a more rigorous 2nd order IMEX Scheme
\citep{Pareschi_Russo05a,Palenzuela_Lehner+09a}. Let us describe here
both of them with reference to a typical system of stiff and non-stiff equations, for the conserved variables
$\bmath{U},\bmath{W} $:
\begin{align}
&\partial_t \bmath{U} =
\bmath{F}(\bmath{U,W}),\nonumber\\   
&\partial_t \bmath{W} = \bmath{G}(\bmath{U,W})+\frac{1}{\eta}\bmath{R}(\bmath{U,W}).   
\end{align}

In the simple 1st-2nd scheme, the explicit part is solved using a modified Eulerian scheme which is second
order in time, while a simple first-order scheme is used to solve  the
implicit one.
\begin{align}
&\bmath{U}^{(1)} = \bmath{U}^n + \Delta t  \bmath{F}(\bmath{U}^n,\bmath{W}^n)/2 ,\nonumber\\
&\bmath{W}^{(1)} = \bmath{W}^n +\Delta t  \bmath{G}(\bmath{U}^n,\bmath{W}^n)/2 +
\frac{\Delta t }{2\eta}\bmath{R}(\bmath{U}^{(1)},\bmath{W}^{(1)}),\nonumber\\
&\bmath{U}^{n+1} = \bmath{U}^n + \Delta t
\bmath{F}(\bmath{U}^{(1)},\bmath{W}^{(1)})\\
&\bmath{W}^{n+1} = \bmath{W}^n +\Delta t  \bmath{G}(\bmath{U}^{(1)},\bmath{W}^{(1)}) +
\frac{\Delta t }{\eta}\bmath{R}(\bmath{U}^{n+1},\bmath{W}^{n+1}),\nonumber
\end{align}

where the implicit step acts only on the electric field $\bmath{E}$,
it is peformed during the inversion from conserved to primitive
variables (see above), accoding to the solution provided in the
previous section.

This symple scheme has the advantage of being easily implementable,
requiring only a modification in the definition of the electric field,
within the algorithm that derives the primitive from the conserved
variables. Moreover the algorithm is well behaved in the case $\eta=0$
[it reduces to solving $ \bmath{R}(\bmath{U,W})   =0$], and thus it
can handle also the
Ideal MHD regime.

The second order IMEX that we have implemented is the following:

\begin{align}
&\bmath{U}^{(1)} = \bmath{U}^n,\nonumber\\
&\bmath{W}^{(1)} = \bmath{W}^n +
\frac{\Delta t \mu}{\eta}\bmath{R}(\bmath{U}^{(1)},\bmath{W}^{(1)}),\nonumber\\
&\bmath{U}^{(2)} = \bmath{U}^n + \Delta t
\bmath{F}(\bmath{U}^{(1)},\bmath{W}^{(1)}),\nonumber\\
&\bmath{W}^{(2)} = \bmath{W}^n + \Delta t
\bmath{G}(\bmath{U}^{(1)},\bmath{W}^{(1)}) + \nonumber\\
&\;\; + \frac{\Delta
  t}{\eta}[(1-2\mu) \bmath{R}(\bmath{U}^{(1)},\bmath{W}^{(1)}) +\mu
\bmath{R}(\bmath{U}^{(2)},\bmath{W}^{(2)})],\nonumber\\
&\bmath{U}^{n+1} = \bmath{U}^n + \frac{\Delta t}{2}[
\bmath{F}(\bmath{U}^{(1)},\bmath{W}^{(1)})+\bmath{F}(\bmath{U}^{(2)},\bmath{W}^{(2)})],\nonumber\\
&\bmath{W}^{n+1} = \bmath{W}^n + \frac{\Delta t}{2}[\bmath{G}(\bmath{U}^{(1)},\bmath{W}^{(1)})+\bmath{G}(\bmath{U}^{(2)},\bmath{W}^{(2)})]+\nonumber\\
&\;\; + \frac{\Delta t}{2\eta}[\bmath{R}(\bmath{U}^{(1)},\bmath{W}^{(1)})+\bmath{R}(\bmath{U}^{(2)},\bmath{W}^{(2)})],
\end{align}

where $\mu=1-1/\sqrt{2}$.

Again the implicit step acts only on the electric field $\bmath{E}$,
it is peformed during the inversion from conserved to primitive
variables. Note however that now the $Q^i$ are no more given
simply by Eq.~\ref{eq:qi}.

However, in comparison with the previous 1st-2nd scheme, this IMEX
scheme requires 3 different steps instead of 2, the steps do not have
the same functional form, and, because of the last step, it is not
well behaved in the case $\eta=0$, and so it cannot be used for Ideal
MHD problems.

All the test that we present in the following have been repeated both
with out 1st-2nd scheme and with the IMEX scheme. As we will discuss,
depending on the problem or the desired accuracy, our proposed 1st-2nd
scheme might offer and easy alternative to more sophisticated IMEX
implementations.

The value of the timestep $\Delta t$ is chosen in
accordance to the CFL condition, with typical Courant numbers ranging
from 0.1 to 0.5 (smaller values have been adopted in high
resistivity runs to avoid large truncation errors in the solution
of the implicit part, due to the smaller diffusive timescale).

As far as the electromagnetic constraints of Eqs.~\ref{eq:gauss}-\ref{eq:divb}
are concerned, all previously developed schemes for (special) relativistic resistive
MHD \citep{Komissarov07a,Palenzuela_Lehner+09a,Dumbser_Zanotti09a}
adopt a {\it divergence-cleaning} approach \citep{Munz_Omnes+00a,Dedner_Kemm+02a}, 
where an augmented system of equations is
introduced (including the charge conservation law), and where the solenoidal
constrain on the magnetic field and Gauss's theorem are not enforced
but preserved by damping and propagating away any violation arising from
numerical truncation errors. Here instead we choose to use a \emph{fully constrained}
scheme, where the charge $q$ appearing in the equation for $\bmath{E}$
is taken directly from Eq.~\ref{eq:gauss} while the divergence-free condition 
on $\bmath{B}$ is treated with staggered grids via the \emph{Upwind Constrained Transport} 
method \citep{Londrillo_Del-Zanna00a,Londrillo_Del-Zanna04a,Del-Zanna_Bucciantini+03a}.
A benefit of this fully constrained approach is that, neglecting dynamo effects ($\aad=0$)
in the purely resistive case, it is possible
to recover the ideal MHD expression simply by setting $\eta=0$ (see
Eq.~\ref{eq:a10}), while all previous scheme recover the ideal regime only
as a limit for low resistivity $\eta \to 0$. However this also
depends on the time-stepping algorith, as we have discussed previously.

To conclude this section, note that
in the resistive case the maximum wave speed is not
limited by the fast magnetosonic mode and may approach the speed of light
independently of the value of the magnetic field.
The Riemann problem at cell interfaces is thus solved for simplicity
using the maximally diffusive (global) Lax-Friedrichs scheme, where 
the fastest characteristic speed is set to be equal to the speed of light.  
The use of such a scheme, as opposed to more
accurate Riemann solvers like the HLL or HLLD, might lead to less
satisfactory results in the ideal MHD limit, where sharp discontinuities
are allow to arise. In the resistive case, where smooth profiles are
expected, the use of more diffusive algorithms is less problematic,
especially in conjunction with high-order reconstruction algorithms, 
a distinguishing feature of our ECHO scheme. Again, we leave the
implementation of more accurate Riemann solvers as a future upgrade.

\section{Test problems}
\label{sec:res} 

In the following we present a set of standard tests. The first three
are done in the purely resistive regime, for comparison with results
previously presented in the literature. Then there are three dynamo problems
in the so-called kinematic regime, and finally a fully dynamical problem
All but the last of these tests are done in a flat, stationary metric,
since they are aimed at evaluating the implementation of the
resistive/dynamo closure, and for a more straightforward comparison with
previous results, and with analytical solutions.

\subsection{Resistive tests}

We present here a set of three resistive tests ($\aad=0$) in a stationary
Cartesian grid (Minkowskian spacetime), to be compared with both
analytical and previously published results. In the following we use a third-order
CENO spatial reconstruction with MC limiter, and both our 1st-order implicit
/ 2nd-order explicit temporal evolution scheme, and the 2nd order IMEX
scheme.  A maximally diffusive
global Lax-Friedrichs Riemann solver (LF)  is employed for upwinding.

\begin{figure}
\resizebox{\hsize}{!}{\includegraphics{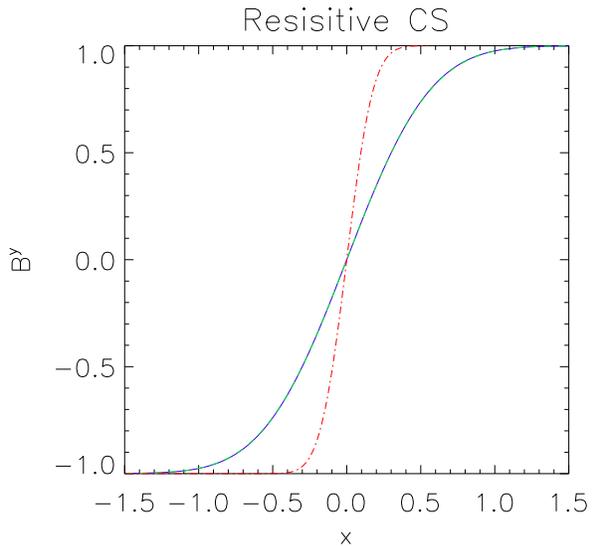}}
\caption{Evolution of the magnetic field in a self-similar current sheet, for
  $\eta=0.01$. The red dot-dashed line is the initial condition at
  $t=1$. The blue solid line is the numerical solution at $t=10$, indistinguishable
  from the green dashed line representing the exact solution Eq.~\ref{eq:currentsheet}.}
\label{fig:currentsheet}
\end{figure}

\begin{figure}
\resizebox{\hsize}{!}{\includegraphics{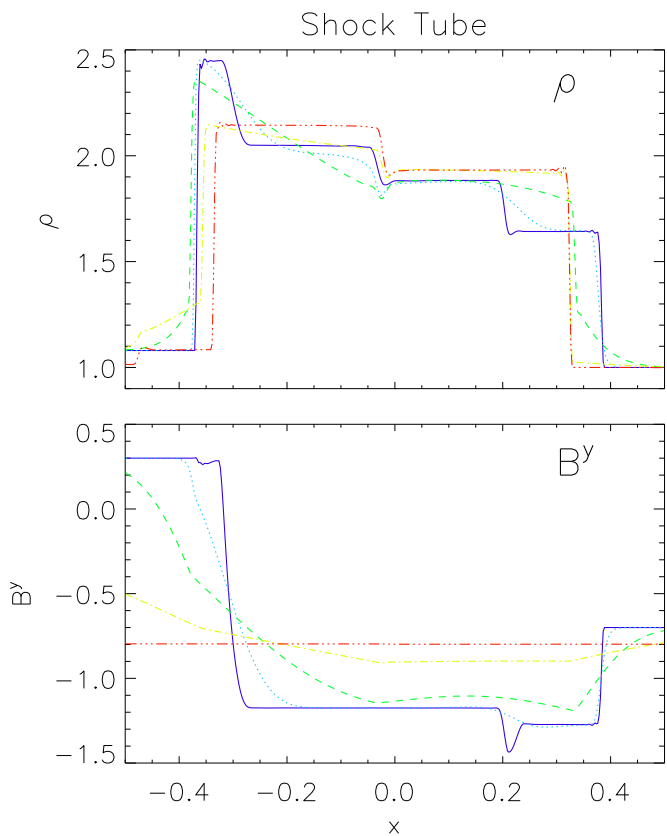}}
\caption{Resistive shock tube problem. The upper panel shows the density
  and the lower panel shows the $y$-component of the magnetic field at
  the final time  $t=0.55$. The blue solid line is the case $\eta=0$,
  the dotted cyan line is the case $\eta=0.01$, the green dashed line
  is for $\eta=0.1$, the yellow dot-dashed line for $\eta=1$, while
  the red dash-triple dotted line refers to $\eta=1000$.}
\label{fig:tube}
\end{figure}

\subsubsection{Self-similar current sheet}
\label{sec:subsubcs}

This problem was first proposed by \citet{Komissarov07a},  and it has
been presented also by \citet{Palenzuela_Lehner+09a} and
\citet{Dumbser_Zanotti09a}. It is a truly resistive problem, which
allows for the analytical self-similar (in $t/x^2$) solution in
the limit of infinite pressure:
\begin{align}
B^y(x,t)=B_0{\rm erf} \bigg(\frac{x}{2\sqrt{\eta t}}\bigg)
\label{eq:currentsheet}
\end{align}
where erf is the error function.

Despite the evolution being almost purely resistive (in principle, in the
limit of infinite pressure only the magnetic field evolves and only the induction
equation needs to be solved), the problem is followed
in the fully dynamical regime. The initial conditions at $t=1$ are:
$\rho=1$, $p=50$, $E^i=0$, $v^i=0$,
$B^x=B^z=0$, $B^y=B^y(x,1)$, with $B_0=1$, and we have adopted an
adiabatic coefficient $\gamma=4/3$. 
Our computational domain extends in the range $x=[-1.5, 1.5]$, and is
covered by a uniform grid with 200 cells. The problem is evolved to
the final time $t=10$. In Fig.~\ref{fig:currentsheet} we compare our
numerical results with Eq.~\ref{eq:currentsheet}, in the case
$\eta=0.01$. Errors and
convergence estimates are presented  in Sect.~\ref{sec:errors}
together with a comparison between the 1st/2nd scheme and the IMEX scheme.

\subsubsection{Resistive shock tube}

This problem was proposed by \citet{Dumbser_Zanotti09a}, and differs
from the one presented in \citet{Palenzuela_Lehner+09a},
which is done in the \emph{transverse} MHD regime. Shock tubes are excellent tests for
monitoring the shock-capturing properties of a numerical scheme. 
The initial conditions are:
\begin{align}
&(\rho,p,v^x,v^y,v^z,B^x,B^y,B^z)=\nonumber\\
&(1.08,0.95,0.4,0.3,0.2,2.0,0.3,0.3)\; {\rm for}\; x<0. 
\end{align}
and
\begin{align}
&(\rho,p,v^x,v^y,v^z,B^x,B^y,B^z)=\nonumber\\
&(1.0,1.0,-0.45,-0.2,0.2,2.0,-0.7,0.5)\; {\rm for}\; x>0. 
\end{align}
and the problem is followed to a final time $t=0.55$. The initial
electric field is set equal to the Ideal MHD value
$E^i=-\epsilon^{ijk}B_jv_k$. The
computational domain extends in the range $x=[-0.5, 0.5]$ and is
covered by a uniform grid with 400 cells. The test was repeated with
the following values for the resistivity: $\eta=0, 0.01, 0.1, 1,
1000$, where the last value was chosen to be big enough such that the
evolution practically corresponds to the zero conductivity case. The adiabatic
coefficient is here $\gamma=5/3$. Results are shown in Fig.~\ref{fig:tube}.

\subsubsection{Resistive rotor}
This is a fully multidimensional problem. The relativistic Ideal-MHD
version was first proposed by \citet{Del-Zanna_Bucciantini+03b}, while
the resistive version has been presented by
\citet{Dumbser_Zanotti09a}.

\begin{figure*}
\resizebox{\hsize}{!}{\includegraphics{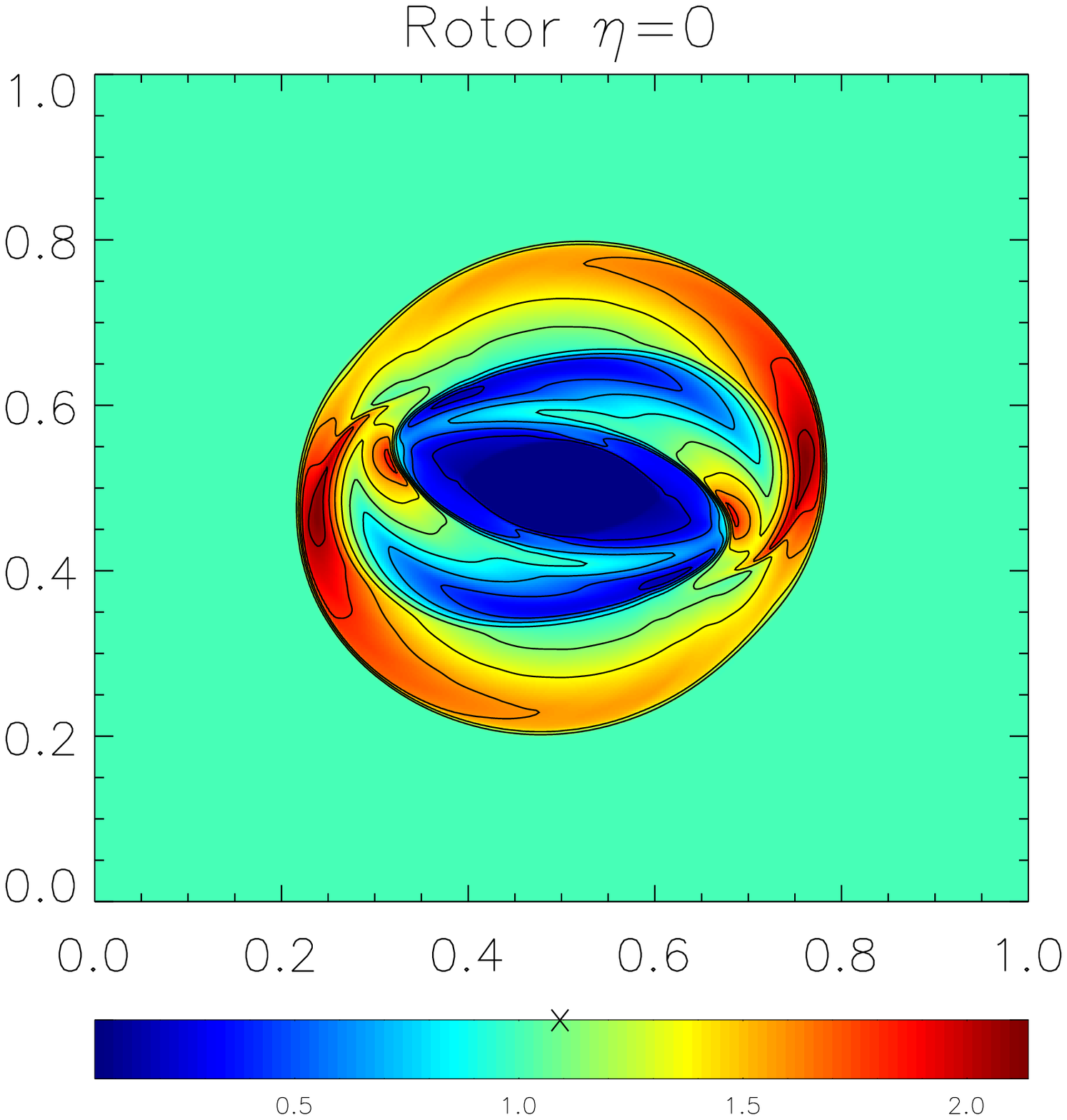}\includegraphics{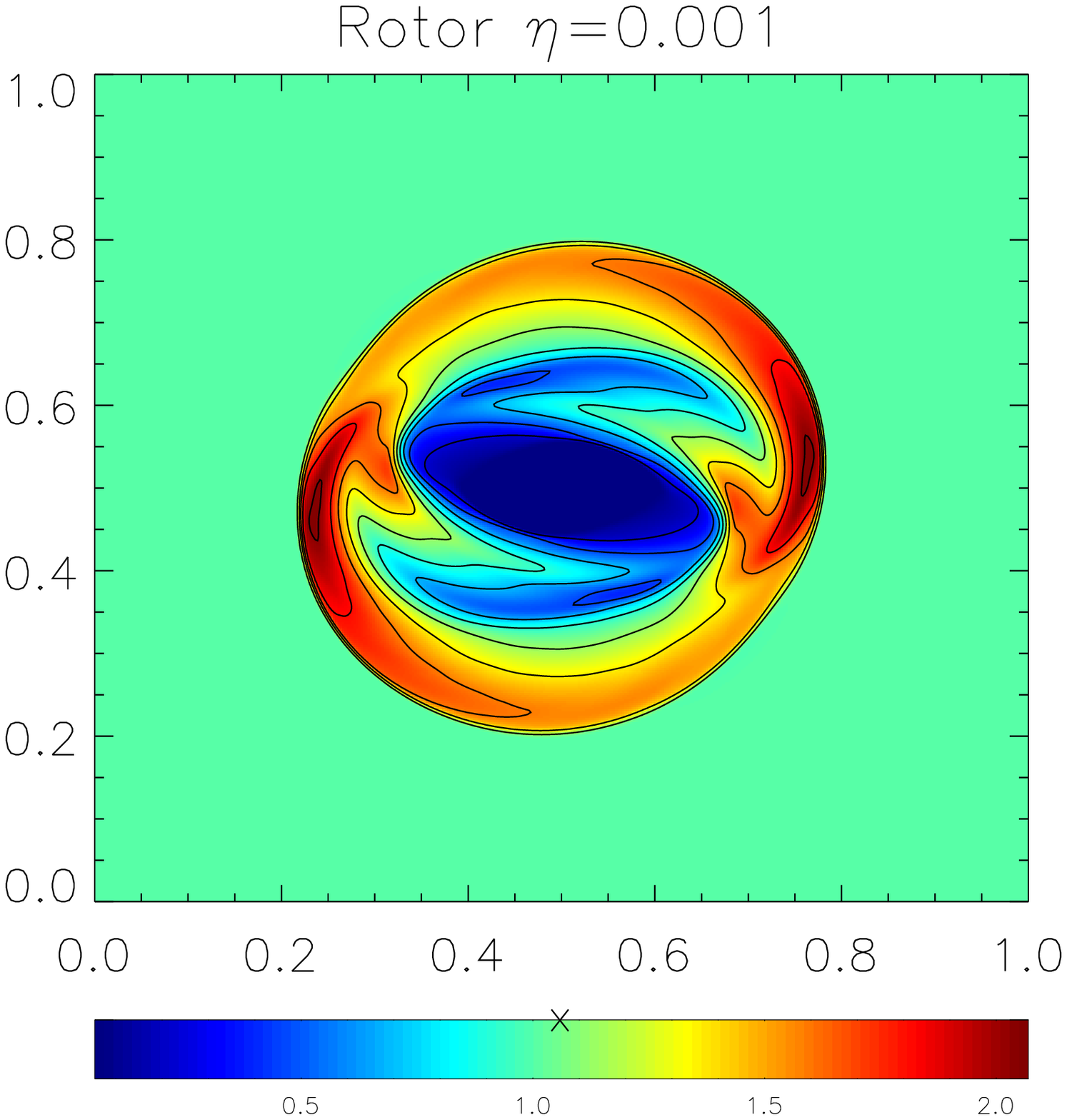}\includegraphics{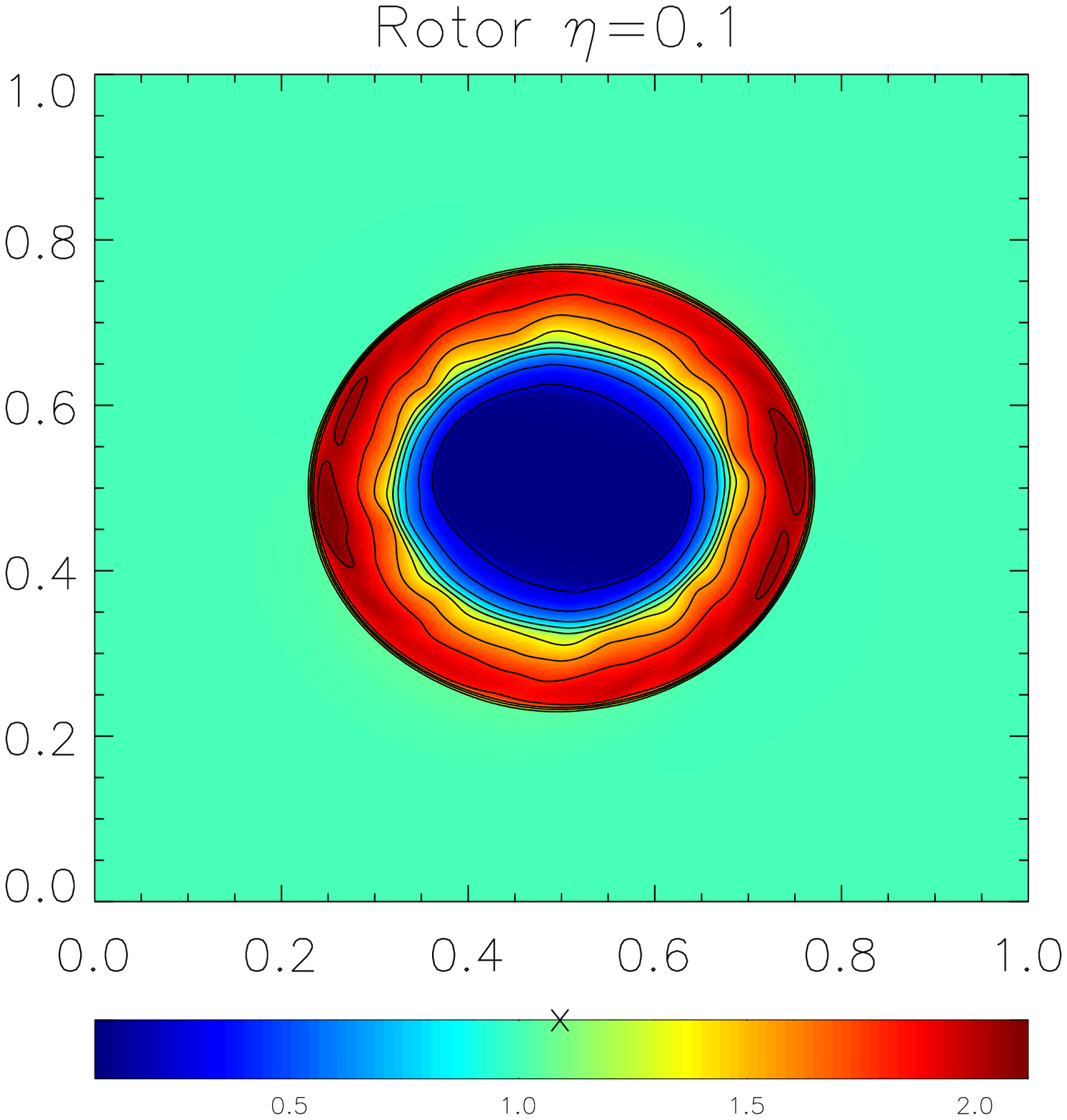}}
\resizebox{\hsize}{!}{\includegraphics{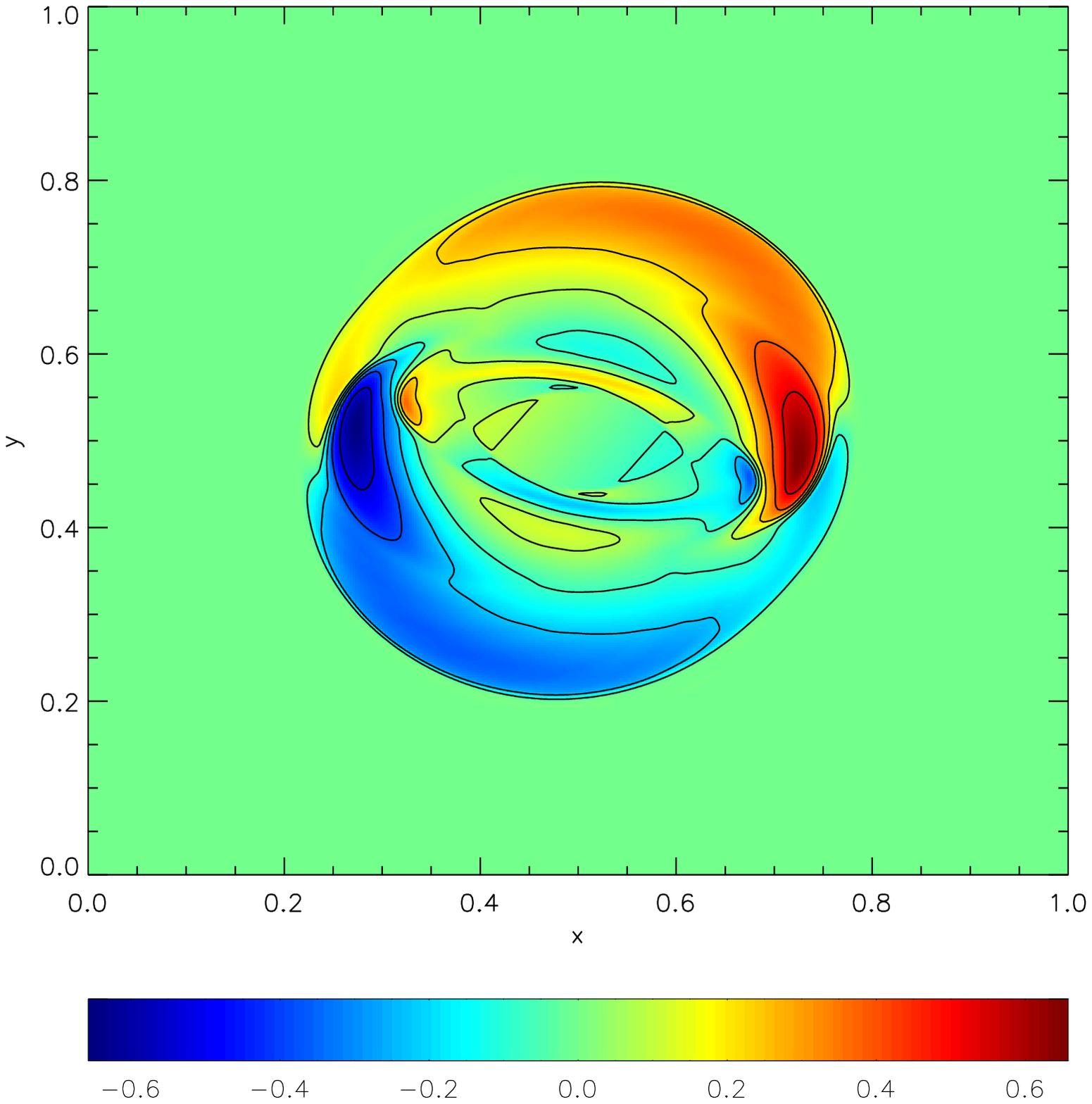}\includegraphics{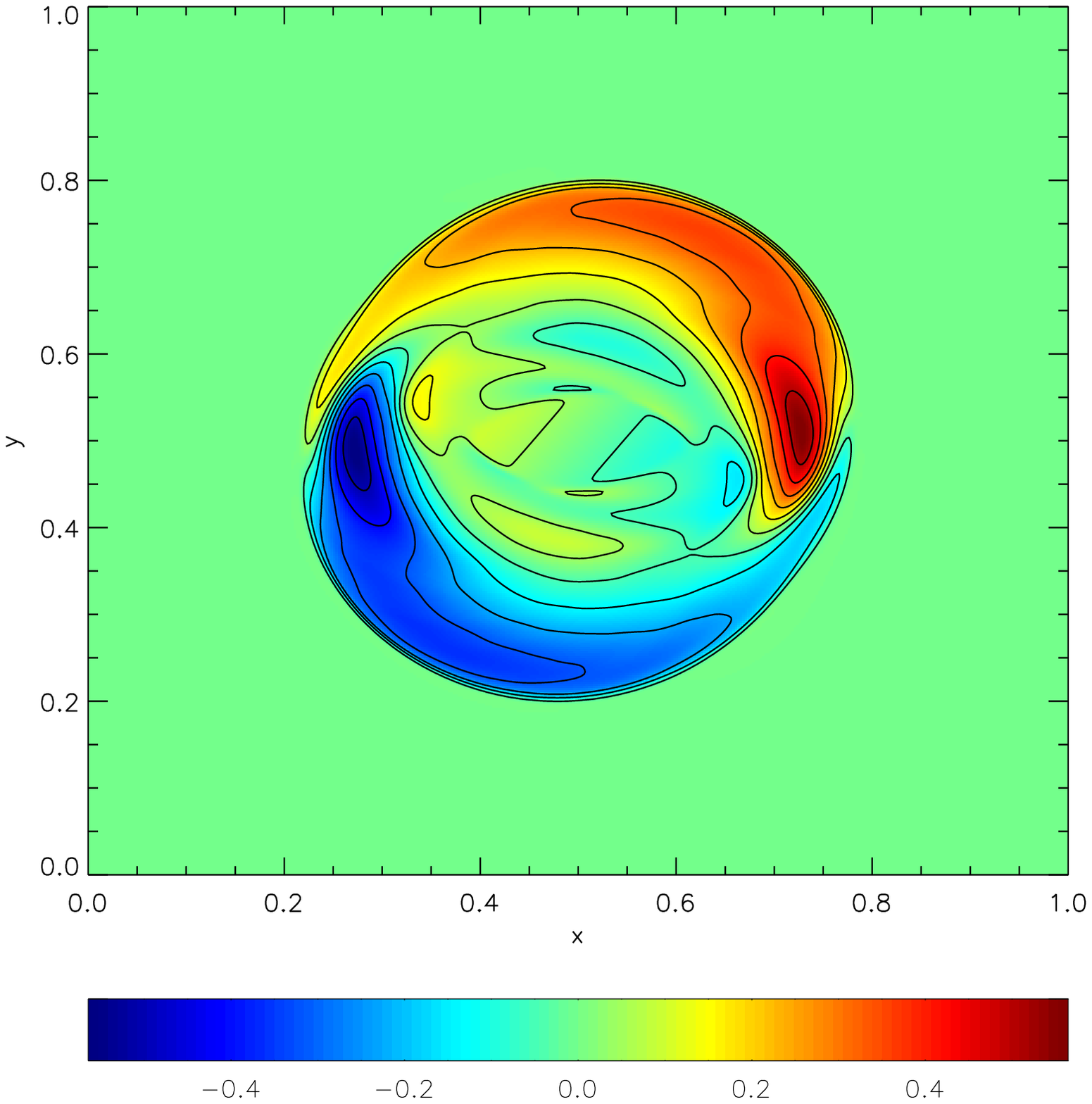}\includegraphics{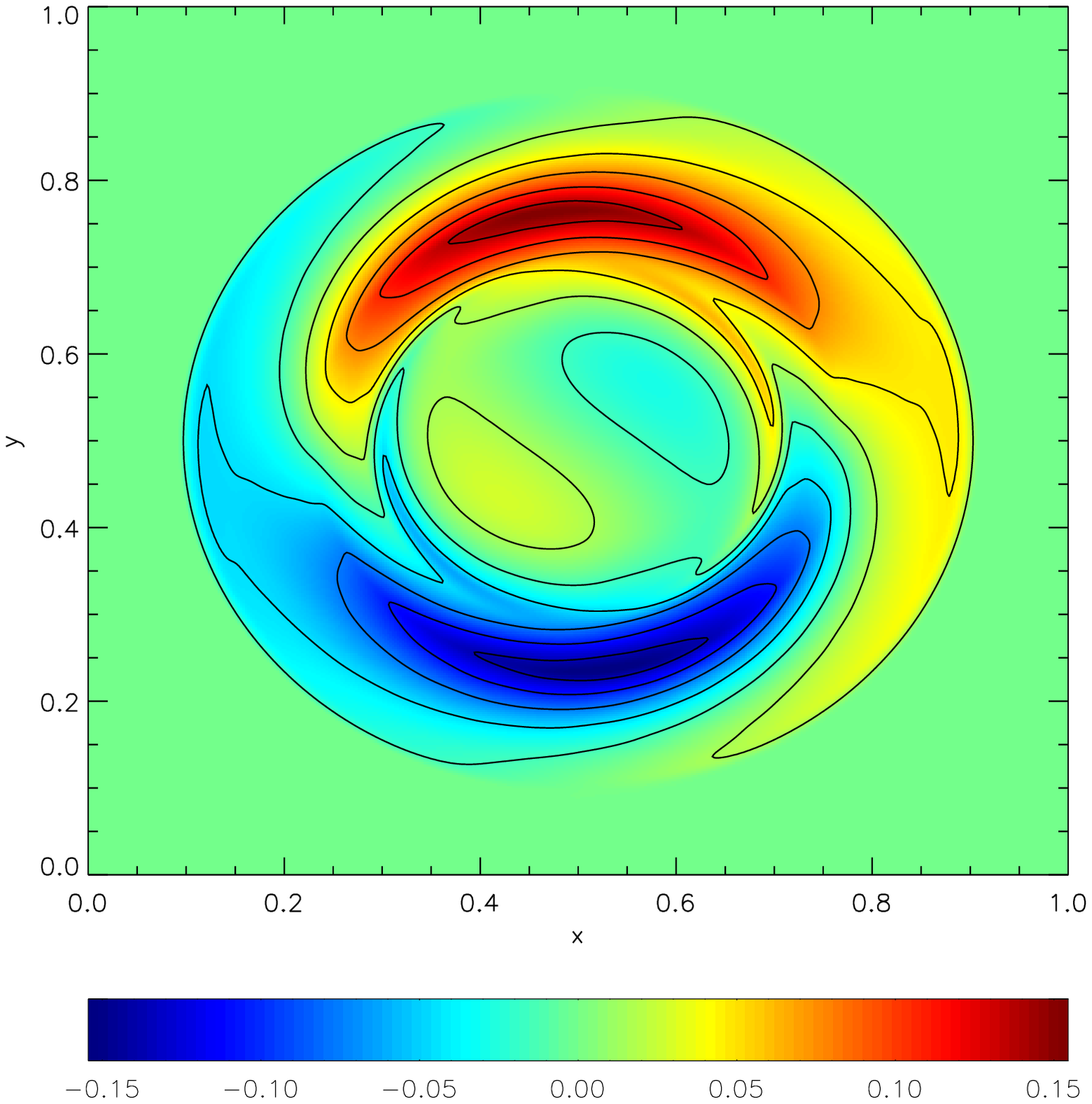}}
\caption{Relativistic rotor. The upper panels show the pressure, while
the lower panels show the z-component of the electric field at the
final time $t=0.3$. Left column: the Ideal case $\eta=0$. Central
column: quasi-ideal case $\eta=0.001$. Right column: resistive case $\eta=0.1$.}
\label{fig:rotor}
\end{figure*}

A circular region with radius $r=0.1$, uniform density $\rho=10$,
and rotating with a uniform angular velocity $\Omega=8.5$ is
located within a medium at rest, with a lower density $\rho=1$. 
The pressure  $p=1$ and the magnetic field $(B^x,B^y,B^z)=(1,0,0)$
are uniform in the whole domain. The adiabatic coefficient is 
$\gamma=4/3$ and the system is followed to a final time $t=0.3$.
The initial electric field is set equal to the ideal MHD value
$E^i=-\epsilon^{ijk}B_jv_k$. The computational domain is
$x=[-0.5,0.5]$, $y=[-0.5,0.5]$, with the rotating region located at
the center, and is covered with a uniform grid of
$400\times 400$ cells. 
The problem is solved in the ideal MHD regime $\eta=0$, in the
quasi-ideal regime $\eta=0.001$, and in the resistive regime
$\eta=0.1$, and the results are shown in Fig.~\ref{fig:rotor}. 
Comparison between the case $\eta=0$ and $\eta =0.001$ suggests that, 
at this resolution,  the intrinsic resistivity of the scheme
is smaller than $\eta=0.001$.
Moreover the case $\eta=0$, when repeated with a HLL solver (not shown here) does not
show any significant improvement with respect to the same case done with LF.

\subsection{Dynamo tests}

Here we present the tests done using the closure Eq.~\ref{eq:dynrel}, leading
to the full version in Eq.~\ref{eq:stiff} of the evolution equation for $\bmath{E}$.
The first three are performed in the so-called kinematic regime, where
only the induction equation is solved and only the electric and
magnetic fields are allowed to change in time. Given the physical nature
of mean-field dynamos, related to small-scale turbulence, the
kinematic approximation is actually commonly adopted as a suitable approximation. 
In fact, the magnetic field
generated by turbulent dynamo action is often well below
equipartition, with respect to the internal energy density, and as
such of negligible dynamical consequences. Moreover, kinematic dynamos
often allow for simple analytical solutions, whereas the non-linear
feedback on the flow can only be dealt with using numerical methods.
As for the resistive cases,  in all of the following we employ a third-order
CENO spatial reconstruction with MC limiter, the maximally diffusive Lax-Friedrichs
solver, and both the usual 1st-order  implicit / 2nd-order explicit
and IMEX
time-stepping schemes. 

\subsubsection{1D steady dynamo}
\label{sec:dyn1d}

This is a very simple test describing the growth of magnetic field in
a stationary medium. It is the dynamo equivalent of the resistive
current sheet presented in section \ref{sec:subsubcs}, because it
admits an analytical solution (see Appendix \ref{app:dynanal}).

We perform a few runs with different  resistivity  $\eta=0.05, 0.1, 0.25$, 
different values of the wave number $k$, while $\aad=0.5$ is kept fixed. 
The computational domain is $x=[-\pi,\pi]$ is covered with 200
uniformly spaced zones, and the problem is followed to a
final time $t=100$. Our initial conditions are chosen to correspond to the growing
eigenmodes of the problem (see Appendix~\ref{app:dynanal}):
\begin{align}
&B^y=0.1\sin{(kx)}, \; B^x=0,  \nonumber\\
&B^z=-0.1\cos{(kx)}
\end{align}

\begin{figure}
\resizebox{\hsize}{!}{\includegraphics{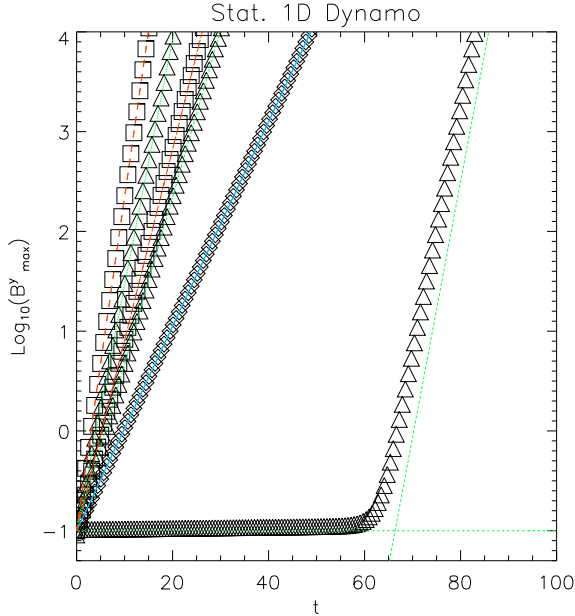}}
\caption{1D steady dynamo. Triangles represent cases with $\eta=0.1$
  (and $k=1,2,5$), squares
  cases with $\eta=0.05$ ($k=1,2$), and diamonds with $\eta=0.25$
  ($k=1$). Overplotted is the expected analytical solution with an
  exponential growth rate $I\omega$: the green
  dotted lines
  refers to cases with  $\eta=0.1$, corresponding
  to $I\omega=0.385,0.567,0.59$; the red dashed lines to  cases with  $\eta=0.05$, corresponding
  to $I\omega=0.44,0.77$; and the solid blue line to $\eta=0.25$,
  corresponding to $I\omega=0.236$.}
\label{fig:1ddynam}
\end{figure}

Fig.~\ref{fig:1ddynam} shows the evolution of the $B^y$ component of
the magnetic field, compared with the corresponding analytical expectations, for
various values of $\eta$ and $k$. It is interesting to note that in
the case $\eta=0.1$ and $k=5$, the magnetic field is expected to remain
unchanged, due to the opposite effects of resistivity and
dynamo. Given the nature of the dynamo closure in Eq.~\ref{eq:dynrel}, due to
round-off and interpolation errors, the evolution of the fastest growing
mode however dominates  asymptotically. This is what is observed  around $t=60$: the
field starts to grow exponentially, with a growth rate corresponding
to the fastest growing mode. The exact time, when this transition
happens, depends on  roundoff and interpolation errors, and differs
for the IMEX scheme.

\begin{figure}
\resizebox{\hsize}{!}{\includegraphics{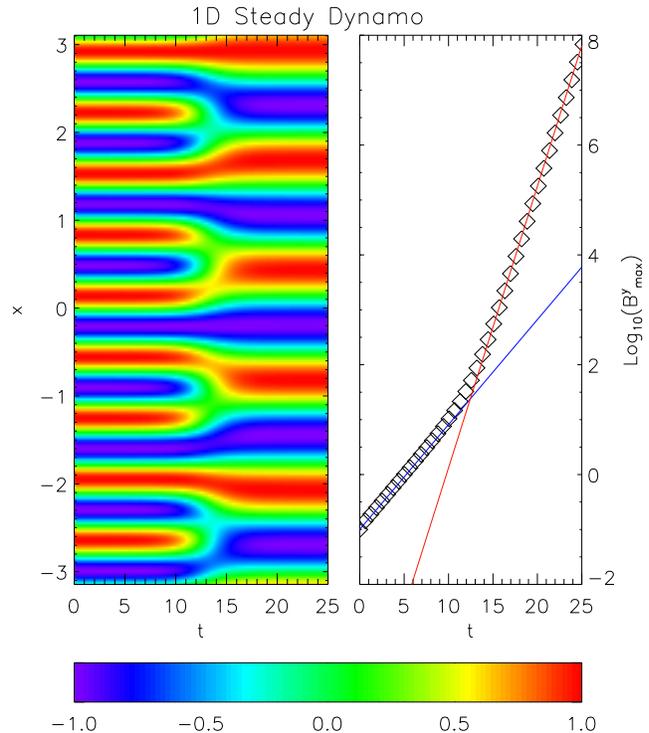}}
\caption{1D steady dynamo, transition to the fastest growing mode. The
  left panel shows the $B^y$ component of the magnetic field,
  normalized to its maximum value. The transition from the $k=9$ mode
to the $k=5$ mode happens around $t=12$. The right panel shows the
evolution of the maximum value of $B^y$. The solid blue line is the
expected growth for a $k=9$ mode, the solid red line is the expected
growth for the fastest $k=5$ mode.}
\label{fig:1ddynevol}
\end{figure}

In Fig.~\ref{fig:1ddynevol} this property of the dynamo solution is
shown for $\eta=0.05$: the  initial solution with a wave number $k=9$ evolves
with a slow growth rate until the fastest growing mode, corresponding to
$k=5$, emerges at $t=12$, to dominate the subsequent evolution. Note that,
given the exponential nature of the solution, the transition is very
sharp.
 Errors and
convergence estimates are presented  in Sect.~\ref{sec:errors}
together with a comparison between the 1st/2nd scheme and the IMEX scheme.

\subsubsection{Thin shear layer}
\label{sec:dynthin}

In this section we present a two-dimensional shear dynamo problem in
the so-called thin-layer approximation, where one of the dimension is
assumed to be much smaller than the other, such that higher order
variations of all quantities in that direction can be neglected.  This
is the relativistic equivalent of the 1D problem investigated by
\citet{Arlt_Rudiger99a}, where spatial variations of the dynamo
coefficient and quenching were also included. 
It is immediately evident the difference between the
non-relativistic and the relativistic case. In the former, one just
need to add a shear term to the induction equation, proportional to the
derivative of the shear velocity along the neglected
dimension. However, this is not possible within our relativistic
formalism, where the modification due to the dynamo process does not
appear in the induction equation, which for us retains the general form
given by Maxwell's equations, but instead in the definition of the electric
field via the closure of Eq.~\ref{eq:dynrel}.  For this reason, we
adopt the thin-layer approach that allows one to recover the 1D results in
the limit of a negligible thickness. This test is representative of
typical conditions in accretion disks, in the limit where one models
the differential rotation with a shear term in the radial direction 
and consider modes extending along the vertical direction.

The computational domain is $x=[-\pi,\pi]$, with periodic boundary
conditions, and $z=[-\delta,\delta]$, with boundary conditions where all
quantities are linearly extrapolated. The  resistivity is set  $\eta=
0.1$, while the dynamo term is $\aad=0.2$.  A shear velocity is
imposed to be $v^y=0.9 z$. The problem is followed to a final time $t=24$, 
corresponding to a dynamo wave period. 
In Appendix \ref{app:dynanal} the relativistic solution is derived, and we
provide also the numerical results for the set of parameters of the
present test.
Our initial conditions are chosen to correspond to the expected growing
eigenmode of the problem
\be
B^y=0.212\cos(x-0.662), \; B^x=0,  \; B^z=0.1\sin(x),
\ee
which corresponds to a growth rate $\omega=-0.238I +0.261$.

\begin{figure}
\resizebox{\hsize}{!}{\includegraphics{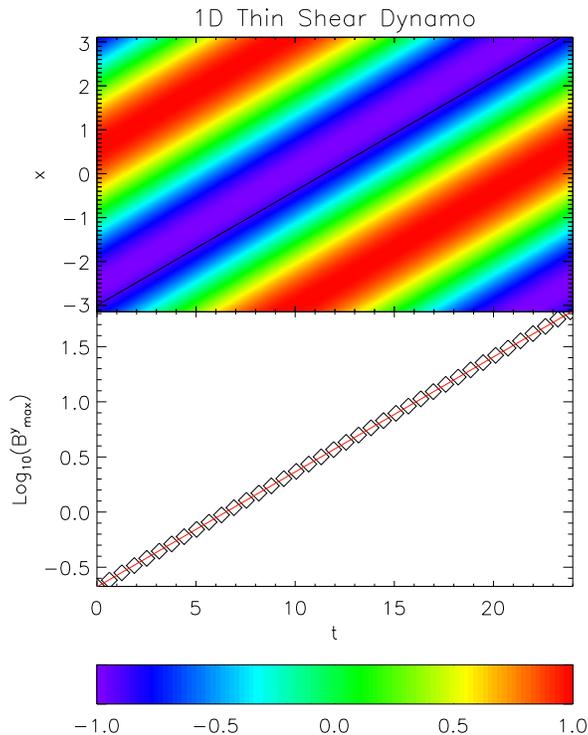}}
\caption{1D thin shear dynamo. The
  upper panel shows the $B^y$ component of the magnetic field,
  normalized to its maximum value, as a function of time and space. The solid
  line represent the analytical phase shift $\propto 0.261t$. The lower panel shows the
evolution of the maximum value of $B^y$. The red solid line is the
analytical solution.}
\label{fig:1dthin}
\end{figure}

The computational domain has 200 equally spaced zones in
the $x$-direction. We have repeated the run varying $\delta$ between
0.05 and 0.2, and the number of cells in the $z$-direction between 5
and 11, and found no appreciable difference in the results. 
In Fig.~\ref{fig:1dthin} we show the evolution and
compare the numerical results with the analytical expectations. 
Both the drifting velocity of the wave and the exponential growth of its
amplitude are properly recovered with an accuracy $\sim 1$\%. Errors and
convergence estimates are presented  in Sect.~\ref{sec:errors}
together with a comparison between the 1st/2dn scheme and the IMEX
scheme, and an estimate of the true growth rate and phase shift.
 
\subsubsection{Kinematic Couette flow}

This is a fully 2D  dynamo test, on a non-Cartesian grid, corresponding
to a Couette flow. We use cylindrical coordinates $R,z$, with a domain
extending in the radial direction in the range $R=[0.1,2]$, with 100
equally spaced zones, and along the $z$-direction in the range
$z=[-1,1]$, with 100 equally spaced zones. This corresponds to an
aspect ratio  $\sim 1$. 

Differential rotation and density (and pressure) stratification are
imposed such that the system is in equilibrium, by assuming a
polytropic relation $p\propto \rho^{4/3}$ and  requiring that the
Bernoulli integral
\be
\ln{(h)}+\frac{1}{2}\ln{(1-v^\phi v_\phi)}-\frac{A}{2}(\Omega-\Omega_0)^2
\ee
is constant in the domain, where $h=1+4p/\rho$ is the specific
enthalpy, $v^\phi=\Omega$ is the angular velocity, $\Omega_0$ is the angular
velocity in $R=0$, and the parameter $A$ is an indicator of the
amount of differential rotation, since
\be
v^\phi v_\phi/(1-v^\phi v_\phi)=\Omega A^2 (\Omega-\Omega_0).
\ee
This approach is equivalent to the one used in models of
differentially rotating NSs
\citep{Komatsu_Eriguchi+89a,Font_Stergioulas+00a,Bucciantini_Del-Zanna11a}.
Here the following values are adopted: $\Omega_0=0.3$,
$\rho(R=0)=p(R=0)=10$, $A^2=5$, which correspond to $v^\phi(R=2)=0.16$.

\begin{figure}
\resizebox{\hsize}{!}{\includegraphics{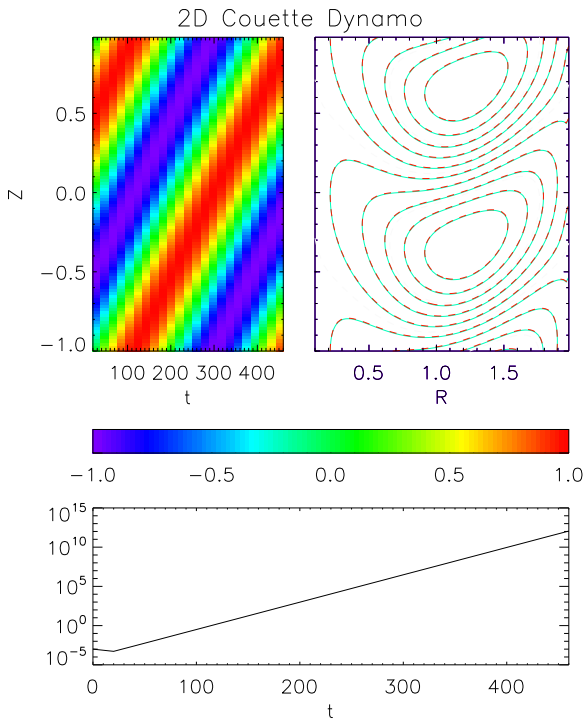}}
\caption{2D Couette flow dynamo. The
  upper-left panel shows the $B^\phi$ component of the magnetic field, in
  $R=1$,
  normalized to its maximum value, as a function of time. The upper
  left panel is a contour plot of $B^\phi$, in the domain at $t=40$
  (solid cyan lines) and
  $t=400$ (dashed red lines), after exactly one wave-cycle.  The lower panel shows the
evolution of the maximum value of $B^\phi$. }
\label{fig:2dcouette}
\end{figure}

The system is followed to a final time $t=460$ corresponding to a
dynamo period. Initially a purely toroidal magnetic field is imposed
in the computational domain
\begin{align}
& B^\phi=0.001 \;{\rm if}\; \sqrt{(r-1)^2+(Z-0.5)^2} < 0.25, \nonumber \\
& B^\phi=-0.001 \;{\rm if}\; \sqrt{(r-1)^2+(Z+0.5)^2} < 0.25,
\end{align}
corresponding to two distinct loops, with a zero net magnetic field in
the domain. We want to remind here that the shape of the initial
magnetic field does not matter, because the system always selects the
fastest growing eigenmode of the dynamo equation. The dynamo 
coefficient is $\aad=0.12$,
while the resistivity is set  to $\eta=0.03$. Periodic boundary conditions
are assumed in the $Z-$direction, while a perfect conductor with $\bmath{E}=0$ is
assumed at the $R=0.1$ and $R=2$ boundaries.
In Fig~\ref{fig:2dcouette} we show the result of the evolution,
focussing on the $B^\phi$ component. It is evident from the top-right
panel that an eigenmode is selected, with a constant drifting speed and
an exponential growth.

\subsubsection{Dynamical NS dynamo in full GRMHD}

As a final test we present here the growth of an $\alpha^2$ dynamo 
(no differential rotation or meridional flow) in a neutron star, in the full 
dynamical general relativistic regime under the XCFC approximation,
as currently employed in the X-ECHO code.
The initial conditions are derived using the XNS
code \citep{Bucciantini_Del-Zanna11a}, to which the reader is referred
for a description of parameters characterizing  the initial equilibrium configuration.
The NS has a central density $\rho_c=1.28\times 10^{-3}$ (in geometrized
units $c=G=M_\odot=1$). The equation of state used for the initial settings
corresponds to the polytropic relation $p=100\rho^2$, however the system is evolved
using an ideal gas EoS as done for all NS models in
\citet{Bucciantini_Del-Zanna11a}. The initial magnetic field is purely
toroidal and its distribution follows the barotropic law:
\begin{align}
\alpha r^2 \sin^2\theta B^\phi = K_m(\alpha^2 r^2 \sin^2\theta
\rho h)^m,
\end{align}
 with magnetic parameters $K_m=10^{-4}$ and $m=1$. $r$ is the radius
 and $\theta$ is the polar angle.

The 2D simulation is performed in spherical-like coordinates for the conformal flat metric
assuming axisymmetry.
The computational domain is $\theta=[-\pi,\pi]$, with reflecting boundary
conditions on the axis, and $r=[0,10]$, with dipole-like reflecting conditions
at the center and zeroth-order extrapolation at the outer radius. The
dipole-like conditions are chosen to allow a dipolar magnetic and
electric field where physical quantities like the radial and
$\theta-$components of the magnetic field do not vanish in $r\to 0$.
The  resistivity is set  $\eta=
0.05$, while the dynamo coefficient is $\aad=0.1$, uniform in the
computational domain.  The problem is followed to a
final time $t=200$. 
The time evolution of the metric is computed once every 100 steps of the
MHD part.  We want to stress here that these values have been
chosen for the sake of having a fast numerical run, and have no physical
significance. The scope here is to perform a test of the code in its
full dynamical regime, and not to address a particular physical problem in
realistic conditions. 
 
\begin{figure}
\resizebox{\hsize}{!}{\includegraphics{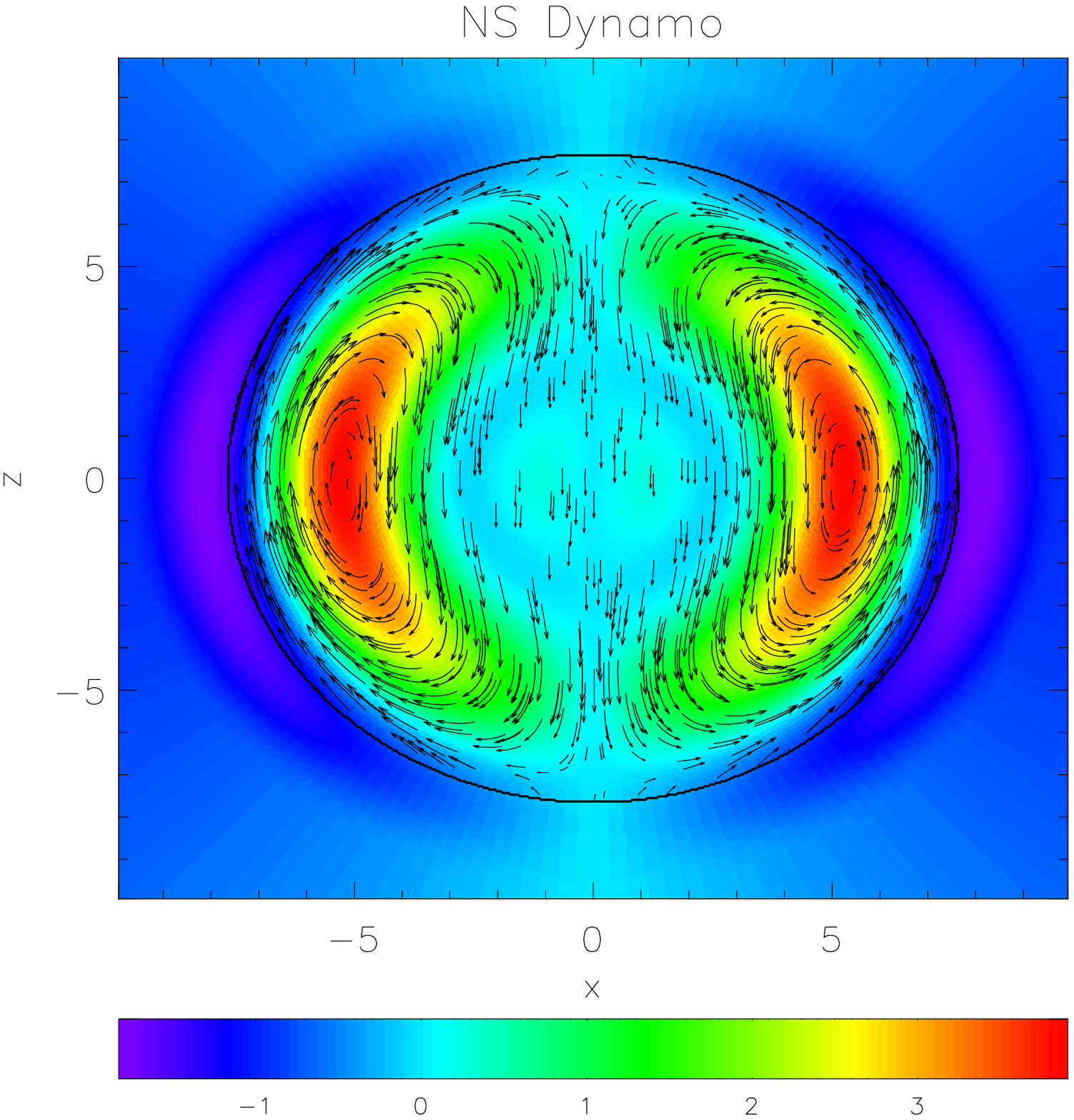}}
\resizebox{\hsize}{!}{\includegraphics{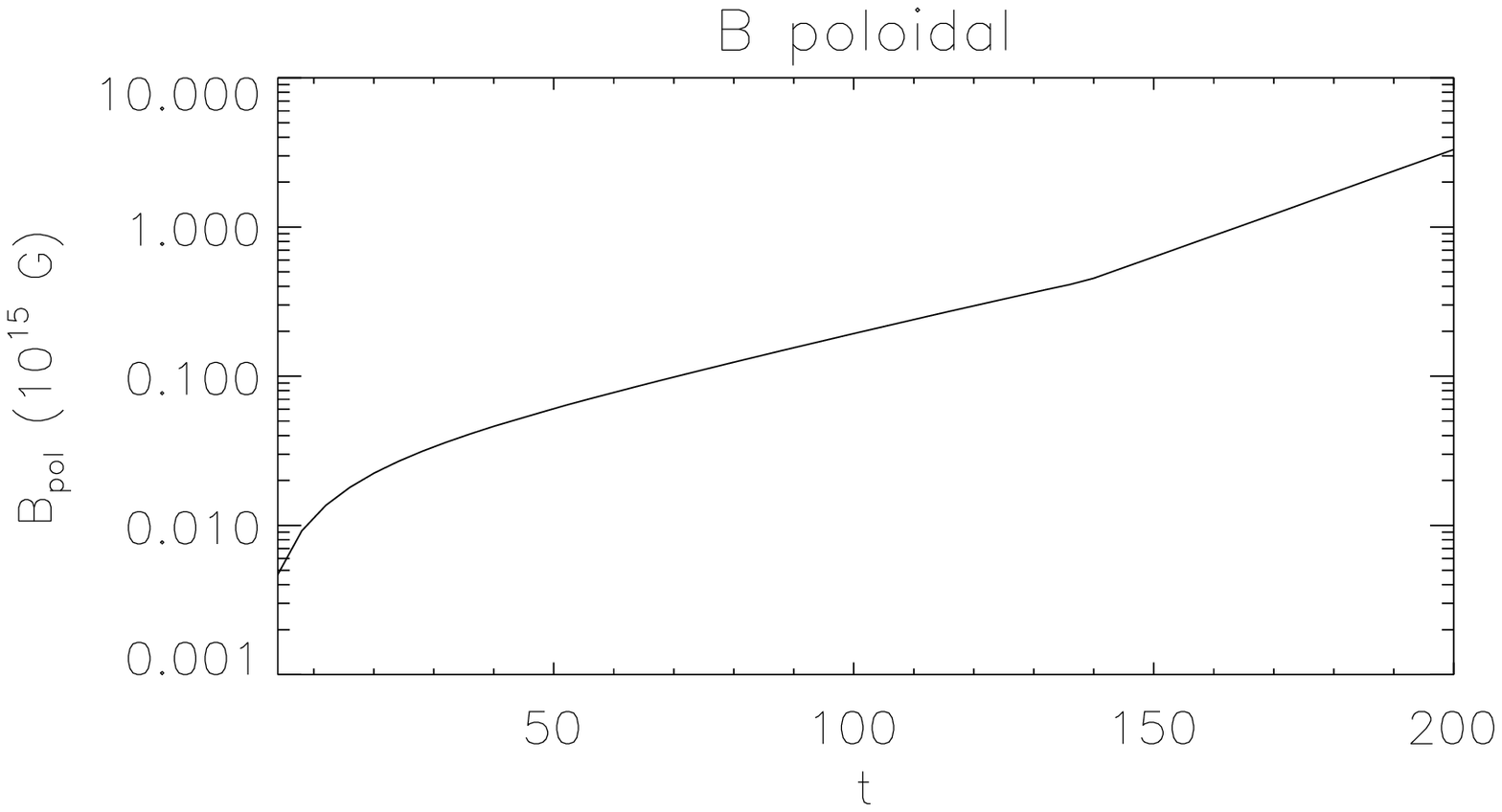}}
\caption{Dynamical NS dynamo. The
  upper-panel shows the poloidal magnetic fieldlines. Colors
  represent the value of the $B^\phi$ component of the magnetic field
  in units $10^{15}$~G. The thick solid line represents the surface of
  the NS. The lower panel shows the
evolution of the maximum value of poloidal magnetic field $\sqrt{B^rB_r+B^{\theta}B_{\theta}}$. }
\label{fig:nsdyn}
\end{figure}

In Fig~\ref{fig:nsdyn} we show the result of the evolution.
The upper panel shows the magnetic configuration that is obtained at
the end of the run. The lower panel shows the growth of the maximum value of the
poloidal magnetic field during the evolution. Having adopted a uniform
$\aad$ term, spurious dynamo action is also present in the atmosphere
surrounding the NS. As a consequence a magnetic field develops and grows in the
atmosphere too, despite it being completely unmagnetized at the
beginning. Again, this is just due to the unphysical choice of a uniform dynamo
term. In the figure, for the sake of clarity, we plot the poloidal
field just inside the star. It is interesting to note that the growth
rapidly approaches an exponential behavior, but that the slope (the growth
rate) changes at around $t=140$. This corresponds to the transition to a
shorter wavelength. In particular, at the beginning the mode that
is excited and grows corresponds to a radial wavelength
of the order of twice the stellar radius (given the initial conditions on
$B^\phi$). At $t=140$ an eigenmode with a shorter radial wavelength of
the order of the stellar radius emerges, corresponding to a faster
growing mode. 

\subsection{Convergence and compartison between the 1st-2nd and IMEX schemes}
\label{sec:errors}
In this section we discuss the convergence properties of our scheme
and compare the different time-stepping approaches. Despite the
simplicity of our 1st-2nd scheme, we do expect that,
depending on the problem, its accuracy might be closer to 1st order,
as opposed to the IMEX which should preserve second order accuracy in
all cases. We want to stress here that the IMEX scheme we have
implemented only acts on the solution of the MHD equations, and it
is not integrated with the metric solver. However in the choice of the
XCFC approach, we already assumed that the metric should be slowly varying
in time with respect to the plasma. In this case we do expect that the
order of the scheme should not be affected by the metric solver.

We want to point
out here that in order to evaluate the convergence of an algorithm,
one needs a test problem with a smooth flow structure at all times
including the initial conditions, and well posed boundary conditions that
preserve the overall accuracy. If a reference
solution, either an analytical solution or a high resolution run (such
that numerical resistivity is smaller than the physical resistivity) for
comparison is not available, one can compute the order of convergence
using relative errors at different resolutions. In Ideal Relativistic MHD such a test was
designed by \citet{Del-Zanna_Bucciantini+03b}, using a large amplitude,
circularly polarized Alfv\'en wave.

\begin{figure}
\resizebox{\hsize}{!}{\includegraphics{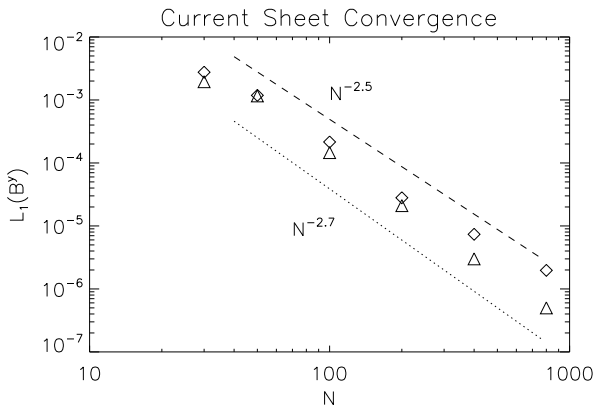}}
\caption{Convergence test for the resistive current sheet
  problem. Relative $L_1$ errors on $B^y$ at $\tilde{t}=10$, with respect to
  a high resolution run with 1600 grid points, in logarithmic
  scale, as function of the number of grid points. Diamonds are for the
  1st-2nd scheme, triangles for the IMEX scheme. Dashed and dotted
  lines reprensent scaling respectively as $-2.5$ and $-2.7$.}
\label{fig:orderresistive}
\end{figure}

In Resistive Relativistic MHD no
such test has been presented yet. The current sheet problem of
Sect.~\ref{sec:subsubcs}, admits an exact solution only in the
limiting case of infinite pressure. For the value of the pressure
that we have used, in line with the previous existing literature,
dynamical effects are present, and they dominate the deviations with
respect to the solution Eq.~\ref{eq:currentsheet}. However, being a one
dimensional problem, it is possible to use a high resolution run as a
reference solution.  In Fig.~\ref{fig:orderresistive} we show 
the relative error on one component of the magnetic field:
\begin{align}
L_1[B^y(t=\tilde{t})]=\frac{1}{N}\sum_{i=1}^N\frac{\| B^y(t=\tilde{t})-B^y_{\rm ref}(t=\tilde{t}) \|}{{\rm max}[B^y (t=\tilde{t})]},
\label{eq:l1norm}
\end{align}
where the sum is done over the value of the quantity in each of the $N$ cells of
the computational domain, and the reference solution at time $\tilde{t}$ is obtained
interpolating a high resolution run with 1600 grid points. It is
interesting to note that in this test problem the IMEX and our 1st-2nd time
stepping algorithm have similar performances, even if the IMEX has a
smaller absolute error. This is because the solution is slowly
evolving in time and the displacement current
is small compared to the conduction current, which is due to spatial
gradients of the magnetic field. For this reason the overall order of
the algorithm is dominated by the order of the explicit part of the solver. Moreover the order seems to be somewhat in
between 2 and 3, as expected from that fact that we use a third order
CENO spatial reconstruction.

\begin{figure}
\resizebox{\hsize}{!}{\includegraphics{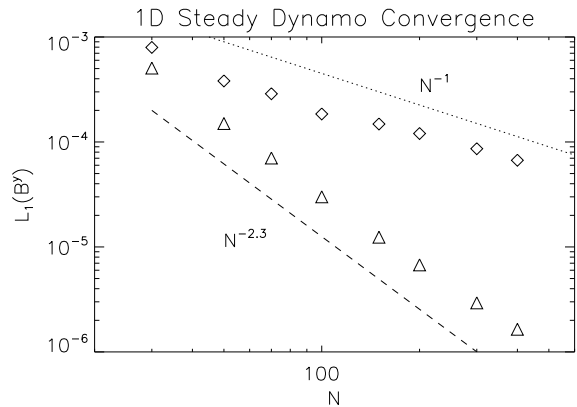}}
\caption{Convergence test for the 1D steady dynamo problem, with
  $k=1$, $\eta=0.1$, $\aad=0.25$. 
 Relative $L_1$ errors on $B^y$ at $\tilde{t}=20$, with respect to
  the analytical solution, in logarithmic
  scale, as function of the number of grid points (per mode wavelength). Diamonds are for the
  1st-2nd scheme, triangles for the IMEX scheme. Dashed and dotted
  lines reprensent scaling respectively as $-2.3$ and $-1$.}
\label{fig:orderdynamo}
\end{figure}

For the relativistic dynamo closure, as opposed to the resistive case,
the 1D steady dynamos admit analytical solutions. The other multidimensional tests we have
presented lack a correct analytical solution and, being
multidimensional, it is
computationally prohibitive to run a high resolution reference case,
however as we will show we can use relative errors at different resolutions.  In Fig. ~\ref{fig:orderdynamo} we show the
 relative error on one component of the magnetic field, defined as
in Eq.~\ref{eq:l1norm} for the 1D steady dynamo  with
  $k=1$, $\eta=0.1$, $\aad=0.25$. These values have been
selected in order for the mode with $k=1$ to be the fastest growing
mode. 1D steady dynamos are rapidly evolving
in time, and with small spatial gradients. The displacement current
dominates over the conduction current. The IMEX scheme performs as
previously, with an order which is again between 2 and 3 for the same
reason as before. The 1st-2nd scheme instead reduces to 1st order as
expected. 

\begin{figure}
\resizebox{\hsize}{!}{\includegraphics{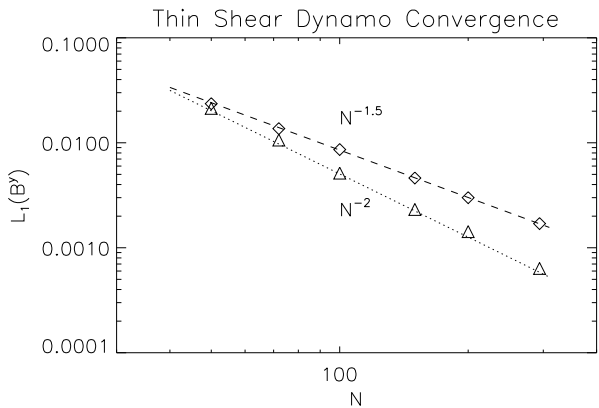}}
\caption{Convergence fit for the thin shear layer dynamo problem.  
Relative $L_1$ errors on $B^y$ at $\tilde{t}=24$, with respect to
  the best fit reference solution, in logarithmic
  scale, as function of the number of grid points (per mode
  wavelength) in the $x-$direction.  Diamonds are for the
  1st-2nd scheme, triangles for the IMEX scheme. Dashed and dotted
  lines reprensent scaling respectively as $-2$ and $-1.5$.}
\label{fig:orderthin}
\end{figure}

We have also attempted to estimate convergence of the solution and
the order of accuracy of our algorithm for the thin shear layer dynamo,
Sect.~\ref{sec:dynthin}. We do not have an analytic solution
available to use as a reference solution, and it is computationally
prohibitive to run a high resolution case, so we proceed comparing
errors in runs a different resolutions. Moreover we do have an
analytic approximated solution and we have an expectation for the
functional form of the eigenmode. In particular we expect the various
quantities to evolve as $e^{-I\omega t}$. So if  we assume that the
unknown true solution changes in time according to this exponential growth,
and that the error of our numerical solutions scales with the number of
grid points $N$ as $N^{-p}$, we can fit simultaneously for the growth rate, phase
shift, and accuracy of the scheme. The result is shown in
Fig.~\ref{fig:orderthin}. This is equivalent to estimate the order of
accuracy by comparing relative errors.
 We find that both for the IMEX
scheme and for our 1st-2nd scheme the best fit estimate for $\omega$ is
$\omega=-I(0.2383 \pm 0.0001) + (0.2626 \pm 0.0005)$, to be compared
with the expectation of the approximated analytical solution
$\omega=-0.238I +0.261$. The best fit for the order gives $p=2\pm 0.1$
for the IMEX and $p=1.5 \pm 0.1$ for the 1st-2nd scheme.
We want to remember here that we use extrapolation for
the boundary conditions in the $z-$direction, and the accuracy of the
solution also depends on the boundary conditions that are used.

In all of our tests, we have verified that, at the resolution
at which they were performed, the relative errors of the 1st-2nd scheme are
 of order a few $10^{-3}$. Unless one requires higher
accuracy, or if the problem involves slowly varying solutions with
strong spatial gradients, or for discontinuous solutions, given its easy implementation, we deem this approach satisfactory. The IMEX scheme in
general shows better convergence already
at 25 points per wavelength of the eigenmode, and should be the
algorithm of choice for more demanding cases.

\section{Conclusion}
\label{sec:con} 

In this paper we have presented a fully covariant mean-field $\alpha-$dynamo
closure of the resistive relativistic MHD equations, and shown how it can be
implemented within a code for numerical $3+1$ General Relativistic MHD.
In particular we have upgraded the \emph{Eulerian Conservative High-Order}
code for GRMHD, in its version for dynamical spacetimes
\citep{Bucciantini_Del-Zanna11a}. 
The X-ECHO scheme employes a fully constrained method for Einstein's equations
based on the \emph{extended conformally flat condition} (XCFC), 
but this particular choice poses no constraint on the applicability of our dynamo
closure, which is unchanged in the Cowling approximation (a static GR metric)
as well as in other formulations of the Einstein equations.
We have shown that our implementation of a dynamo effect  is
straightforward for any numerical scheme that has already been extended
to include resistivity, since the novel generalized Ohm's law is very similar
to that for resistive GRMHD and does not pose any additional numerical difficulty.
This is true in the simple case proposed here, where an isotropic resistivity
and an isotropic $\alpha$-dynamo term (thus simple proportionality between
the electric and magnetic fields in the comoving frame) have been considered.
Far more complex closures have been developed for non-relativistic MHD, 
however it is not obvious if and how fully covariant equivalent formulations 
could be found at all. We stress again that determining
realistic values for the parameters
that are used in the closure, or even finding an appropriate closure, is usually non trivial, and often
requires the use of mesoscale informations, and extrapolation of flow
properties to small unresolved scales. It rests to be proved that a mean
field approach can achieve full resolution of microscopic physics on 
 the macro-scale. 

This is the first numerical implementation of a fully covariant dynamo closure
for relativistic MHD in the general dynamical regime. Instead of starting with the
simpler Minkowskian case, we have directly proposed and applied our model
to full GRMHD, exploiting the $3+1$ formalism, in both static and evolving spacetimes.
We have adopted a fully constrained strategy, to retain the general philosophy
of the ECHO and X-ECHO schemes. The charge density is derived from
Gauss theorem, and it is not evolved as an independent quantity like
in previous formulations (no appreciable differences are observed in the
numerical tests available in the literature).
The solenoidal condition on the magnetic field is preserved to machine
accuracy using the Upwind Constrained Transport method on a staggered grid.
Stiff terms due to small resistivity in the evolution equation for the (spatial)
electric field have been treated both with a simple implicit time-stepping procedure,
which allows us, contrary to previous works, to obtain the ideal case of
a perfectly conducting plasma simply by setting the resistivity coefficient to zero, 
and not just as a limit for small resistivity, and with a more roboust
IMEX scheme. We have shown that depending on the problem, and in
particular on the relative importance of spatial versus temporal
gradients, a simple 1st-order implicit scheme can give satisfactory
results, while the IMEX scheme tends to give more reliable performances
independently of them.

We have presented a set of standard tests, easily implementable and,
where possible, we have compared the numerical results with the analytical
expectations, confirming the robustness of the implementation. The
majority of the tests have been performed in the so-called kinematic
regime. This choice has a physical motivation: mean-field dynamo action
arises from small-scale motions, that are almost always strongly
subsonic. Their kinetic energy is negligible with respect to the
internal or gravitational energy of the system. Dynamo action is
supposed to be quenched once the strength of the magnetic field
reaches equipartition with the kinetic energy of the turbulent
motions. As such, one expects that dynamo amplified magnetic fields
cannot reach values high enough to affect the global dynamics. However,
to verify the stability and robustness of the implementation in a
more demanding regime, which must be fully dynamic and closer to a physical
application, we have also presented a full dynamical case applied to a NS in
GRMHD with a time-dependent metric. 

Our results show that it is possible to implement within codes for
numerical relativity, a closure that in principle can allow one to
 model effects arising from dynamics at small scales, that
would be prohibitive to follow in global simulations. The general idea of
sub-gridding modeling effects, has been widely developed in the context of
classical fluid dynamics, but is still in its infancy in the field of
numerical relativity. There are several outstanding problems of
relativistic fluid dynamics, ranging from the origin of relativistic
engines, to their characterization to the dissipative evolution of
their magnetic fields, that involve extensive dynamical ranges
in space and time. The development of a clever sub-gridding
approach offers an interesting possibility to investigate and model
physical processes that otherwise would require a resolution that
would be computationally prohibitive. This might lead to a different
approach and to a rapid advancement in the field. 

We plan to apply our code for dynamo in GRMHD to both accreting disks
around BHs and in proto-NSs. These two different environments are  
fundamentally related to the more promising engines for GRBs, and
might have important implications for the general modeling of
core-collapse supernovae. Strong magnetic fields have also been
invoked for Short GRBs, which are commonly considered to be a possible
electromagnetic counterpart of binary mergers. Indeed, much of the MHD modeling done until
now has focused on the large-scale properties, but it has been shown
that the expected magnetic configurations can be highly unstable, and
a turbulent cascade is expected. Understanding if and under what
conditions a mean-field dynamo can operate, what is its efficiency,
and the geometry and topology of the resulting field, might help to
put stronger constraints onto the environment within which they are
supposed to operate.

\section*{Acknowledgments}
The authors wish to thank A.~Bonanno, O.~Zanotti, C.~Palenzuela, and A.~Bandenburg,
for fruitful discussions on issues relating dynamos, numerical
relativity, and the solution of the implicit stiff equations. We thank
the referee for his fruitful suggestions and comments.

\vspace{-0.cm}

\appendix
\section{Dynamo solution for a relativistic thin shear layer}
\label{app:dynanal}

We present here an approximated analytical solution for a simple
relativistic shear layer (see problems in
Sect.~\ref{sec:dyn1d}, \ref{sec:dynthin}). It is the relativistic
generalization of a problem presented by \citet{Arlt_Rudiger99a}.
Let us consider a shear layer, with an extension in the $z$-direction much
smaller than in the $x$-direction, where we assume that all quantities
do not depend on $y$. Within this layer the velocity
has the following profile: $v^x=v^z=0$, $v^y=Sz$, where $S$ is the
shear parameter, and does not change in time.

We look for solutions of the form $f(z)\mathrm{e}^{I(\omega t - kx)}$,
where $I=\sqrt{-1}$.
Symmetry tells us that the electric and magnetic fields must have the following parities:
\begin{align}
B^x, E^x \propto  \bigg[\sum_{i=0}^n a_i z^{2i+1}\bigg] e^{I(\omega t -  kx)}, \nonumber\\
B^y, B^z, E^y, E^z \propto  \bigg[\sum_{i=0}^n a_i z^{2i}\bigg],
e^{I(\omega t - kx)}
\end{align} 
where the coefficients $a_i = B^x_i, E^x_i, B^y_i, ...$, complex numbers, differ for
the various quantities. We will look for a solution up to a $z^2$ order.

First, the divergence-free condition on the magnetic field implies
$B^z_1=IkB^x_1/2$.
We start from Eq.~\ref{eq:stiff} and from the induction equation for the
magnetic field, which yield
\begin{align}
&\gamma^{-1/2}\partial_t (\gamma^{1/2}B^i) =
\epsilon^{ijk}\partial_j (\alpha E_k+ \epsilon_{klm}\beta^lB^m).
\end{align}
In the special case of a flat metric in Cartesian coordinates we have
$\alpha=\gamma=1, \beta^i=0$, then
\begin{align}
& \aad \left(2 S \left(E^z_1
   z^2+E^z_0\right)-2 B^x_0\right)+i B^x_0 k S z^2\nonumber\\
& + 4 B^y_1 \eta +2 B^z_0 S+2 I \eta
    E^x_0 \omega+2 E^x_0=0 \\
&\aad \left(2 B^z_0+z^2 (2 E^x_0
   S+I B^x_0 k)\right)\nonumber\\
&-2 I (z^2 (I B^x_0
   S+B^y_1 \eta  k+\eta  E^z_1 \omega-I
   E^z_1) \nonumber\\
&+B^y_0 \eta  k+E^z_0 (\eta 
   \Omega-I))=0\\
&(-2 \aad (S^2 z^2-1)
   \left(B^y_1 z^2+B^y_0\right)-B^x_0 \eta 
   (k^2 z^2-2)\nonumber\\
&-2 (-I B^z_0 \eta  k+z^2
   (-I \eta  E^x_0 k S+i \eta  E^y_1  \omega+2
   \eta E^z_1 S-\nonumber\\
& E^y_1 S^2
   z^2+E^y_1)+E^y_0 (I \eta  \omega-S^2
   z^2+1)))=0\\
&2E^y_1-I B^x_0 \omega=0\\
& (-I E^x_0 + E^z_0 k + B^y_0 \omega + E^z_1 k z^2 + B^y_1 \omega
z^2)=0\\
&(2 I E^y_0 k - 2 I B^z_0 \omega + k (2 I E^y_1 + B^x_0 \omega) z^2)=0.
\end{align}
If we impose these relations to be satisfied both at the
${\cal O}(1)$ and ${\cal O} (z^2)$ order  we can get a set of 4 linear
equations for the variables $B^{x}_{0}, B^y_{0,1} , B^z_{0}$:
\begin{align}
&-I B^x_0 \omega + 
 1/(-I + \eta \omega)^4 (4 a^3 B^z_0 \eta S^3 + 
    B^x_0 \eta (-I + \eta \omega)^2 \nonumber\\
&(2 S^2 + k^2 (1 + I \eta \omega)) + 
    2 a^2 \eta S^2 (B^x_0 (-2 - 2 I \eta \omega) + \nonumber\\
&
       k (-2 I B^y_0 \eta S + B^z_0 (-I + \eta \omega))) + 
    2 I a (-I + \eta \omega) \nonumber\\
&(B^y_1 (1 + 2 I \eta \omega + \eta^2 (4 S^2 - \omega^2)) + 
       \eta S^2 (2 B^z_0 S + \nonumber\\
&B^y_0 (-\eta k^2 - I \omega + \eta \omega^2))))=0\\
&1/(-I + \eta \omega)^2 (-\aad^2 B^z_0 S - 
   I (-I + \eta \omega) (2 B^y_1 \eta - B^y_0 \eta k^2 \nonumber\\
&+ B^z_0 S - I B^y_0 \omega + 
      B^y_0 \eta \omega^2) + 
   \aad (B^x_0 + I B^x_0 \eta \omega + \nonumber\\
&I k (B^z_0 + B^y_0 \eta S + I B^z_0 \eta \omega))=0\\
&B^z_0 \omega + (I k (\aad B^y_0 + \eta (B^x_0 + I B^z_0 k)))/(-I + \eta \omega)=0\\
&B^y_1 \omega + 1/(2 (-I + \eta \omega)^3)
    k (-2 I a^3 B^z_0 S^2 - 2 (B^y_1 \eta k \nonumber\\
&+ I B^x_0 S) (-I + \eta \omega)^2 - 
      2 \aad^2 S (B^y_0 \eta k S + B^x_0 (-I + \eta \omega))
      \nonumber\\
&+ 
      a (-I + \eta \omega) (2 S (2 B^y_1 eta + B^z_0 S) + 
         B^x_0 k (-I + \eta \omega))))=0.
\end{align}
Imposing that the determinant of the matrix representative of this
system vanishes, provides us with a fifth-order equation for $\omega$
as a function of $k, \eta, \aad, S$ which is our dispersion relation.

For example, in the case $\aad=0.2, \eta=0.1, k=1,$ and $S=0.9$ we find a growing mode
solution with $\omega=0.261-0.238 I$. To this mode it corresponds an
eigenvector with $|B^y_0|/|B^x_0|=2.13$ and  with a phase difference
between these two components $\delta \phi = 0.66$.
The non-relativistic case, instead, where displacement currents and
charge densities are neglected, and where an exact analytical solution can
be found, gives  $\omega=0.254-0.240 I$.

In the case $S=0$ it is possible to find an exact analytical solution
for the dispersion relation and the eigenmodes
\begin{align}
I (-\aad k - \eta k^2 - I \omega + \eta \omega^2)/(-I + \eta \omega)=0,
\end{align}
which gives:
\begin{align}
\omega= I[1 \pm \sqrt{1 + 4 \eta k (\aad-\eta k)}]/(2 \eta).
\end{align}
Exponentially growing modes are possible only for $k < \aad/\eta$. The
quantity $\aad/\eta k$ is the dynamo number.  The fastest growing
mode has $k=\aad/(2\eta)$ and a growth rate $\omega_{{\rm max}} =
(\sqrt{1+\aad^2} -1)/(2\eta)$. The eigenmode
corresponding to the growing solution is
\begin{align}
&B^y = IB^z, B^x=0, E^x=0\nonumber\\
&E^y=\frac{I(\aad B^y +IB^z\eta k)}{I-\eta\omega}\\
&E^z=\frac{I\aad B^z +B^y\eta k}{I-\eta\omega}.\nonumber
\end{align}
This solution can be compared with the non-relativistic case, where
the displacement current is neglected
\begin{align}
&\omega=Ik(a-\eta k)\nonumber\\
&B^y = IB^z, B^x=0.\\
&E^y=\aad B^y +IB^z\eta k\nonumber\\
&E^z=\aad B^z -IB^y\eta k.\nonumber
\end{align}


\bibliography{ms2.bib}{}

\begin{thebibliography}{91}
\expandafter\ifx\csname natexlab\endcsname\relax\def\natexlab#1{#1}\fi

\bibitem[{{Anile}(1989)}]{Anile89a}
{Anile} A.~M., 1989, {Relativistic fluids and magneto-fluids: With applications
  in astrophysics and plasma physics}, Anile A.~M., ed.

\bibitem[{{Arlt} \& {R\"{u}diger}(1999)}]{Arlt_Rudiger99a}
{Arlt} R., {R\"{u}diger} G., 1999, \aap, 349, 334

\bibitem[{{Balbus} \& {Hawley}(1998)}]{Balbus_Hawley98a}
{Balbus} S.~A., {Hawley} J.~F., 1998, Reviews of Modern Physics, 70, 1

\bibitem[{{Barkov} \& {Komissarov}(2008)}]{Barkov_Komissarov08a}
{Barkov} M.~V., {Komissarov} S.~S., 2008, \mnras, 385, L28

\bibitem[{{Belvedere} \& {Molteni}(1984)}]{Belvedere_Molteni84a}
{Belvedere} G., {Molteni} D., 1984, Physica Scripta Volume T, 7, 163

\bibitem[{{Benson} \& {Babul}(2009)}]{Benson_Babul09a}
{Benson} A.~J., {Babul} A., 2009, \mnras, 397, 1302

\bibitem[{{Bisnovatyi-Kogan}, {Moiseenko} \&
  {Ardelyan}(2008){Bisnovatyi-Kogan}, {Moiseenko}, \&
  {Ardelyan}}]{Bisnovatyi-Kogan_Moiseenko+08a}
{Bisnovatyi-Kogan} G.~S., {Moiseenko} S.~G., {Ardelyan} N.~V., 2008, Astronomy
  Reports, 52, 997

\bibitem[{{Blackman} \& {Field}(2002)}]{Blackman_Field02a}
{Blackman} E.~G., {Field} G.~B., 2002, Physical Review Letters, 89, 265007

\bibitem[{{Blandford} \& {Payne}(1982)}]{Blandford_Payne82a}
{Blandford} R.~D., {Payne} D.~G., 1982, \mnras, 199, 883

\bibitem[{{Blandford} \& {Znajek}(1977)}]{Blandford_Znajek77a}
{Blandford} R.~D., {Znajek} R.~L., 1977, \mnras, 179, 433

\bibitem[{{Bonanno}, {Rezzolla} \& {Urpin}(2003){Bonanno}, {Rezzolla}, \&
  {Urpin}}]{Bonanno_Rezzolla+03a}
{Bonanno} A., {Rezzolla} L., {Urpin} V., 2003, \aap, 410, L33

\bibitem[{{Braithwaite} \& {Spruit}(2006)}]{Braithwaite_Spruit06a}
{Braithwaite} J., {Spruit} H.~C., 2006, \aap, 450, 1097

\bibitem[{{Brandenburg} \& {Dobler}(2002)}]{Brandenburg_Dobler02a}
{Brandenburg} A., {Dobler} W., 2002, Astronomische Nachrichten, 323, 411

\bibitem[{{Brandenburg} \& {Subramanian}(2005)}]{Brandenburg_Subramanian05a}
{Brandenburg} A., {Subramanian} K., 2005, \physrep, 417, 1

\bibitem[{{Bucciantini} \& {Del Zanna}(2011)}]{Bucciantini_Del-Zanna11a}
{Bucciantini} N., {Del Zanna} L., 2011, \aap, 528, A101+

\bibitem[{{Bucciantini} {et~al}\mbox{.}(2008){Bucciantini}, {Quataert},
  {Arons}, {Metzger}, \& {Thompson}}]{Bucciantini_Quataert+08b}
{Bucciantini} N., {Quataert} E., {Arons} J., {Metzger} B.~D., {Thompson} T.~A.,
  2008, \mnras, 383, L25

\bibitem[{{Bucciantini} {et~al}\mbox{.}(2009){Bucciantini}, {Quataert},
  {Metzger}, {Thompson}, {Arons}, \& {Del Zanna}}]{Bucciantini_Quataert+09a}
{Bucciantini} N., {Quataert} E., {Metzger} B.~D., {Thompson} T.~A., {Arons} J.,
  {Del Zanna} L., 2009, \mnras, 396, 2038

\bibitem[{{Burbidge}, {Layzer} \& {Phillips}(1981){Burbidge}, {Layzer}, \&
  {Phillips}}]{Burbidge_Layzer+81a}
{Burbidge} G., {Layzer} D., {Phillips} J.~G., 1981, \araa, 19

\bibitem[{{Cordero-Carri\'{o}n} {et~al}\mbox{.}(2009){Cordero-Carri\'{o}n},
  {Cerd\'{a}-Dur\'{a}n}, {Dimmelmeier}, {Jaramillo}, {Novak}, \&
  {Gourgoulhon}}]{Cordero-Carrion_Cerda-Duran+09a}
{Cordero-Carri\'{o}n} I., {Cerd\'{a}-Dur\'{a}n} P., {Dimmelmeier} H.,
  {Jaramillo} J.~L., {Novak} J., {Gourgoulhon} E., 2009, \prd, 79, 024017

\bibitem[{{Dedner} {et~al}\mbox{.}(2002){Dedner}, {Kemm}, {Kr\"{o}ner}, {Munz},
  {Schnitzer}, \& {Wesenberg}}]{Dedner_Kemm+02a}
{Dedner} A., {Kemm} F., {Kr\"{o}ner} D., {Munz} C.-D., {Schnitzer} T.,
  {Wesenberg} M., 2002, Journal of Computational Physics, 175, 645

\bibitem[{{Del Zanna}, {Bucciantini} \& {Londrillo}(2003{\natexlab{a}}){Del
  Zanna}, {Bucciantini}, \& {Londrillo}}]{Del-Zanna_Bucciantini+03a}
{Del Zanna} L., {Bucciantini} N., {Londrillo} P., 2003{\natexlab{a}}, Memorie
  della Societa Astronomica Italiana Supplementi, 1, 165

\bibitem[{{Del Zanna}, {Bucciantini} \& {Londrillo}(2003{\natexlab{b}}){Del
  Zanna}, {Bucciantini}, \& {Londrillo}}]{Del-Zanna_Bucciantini+03b}
---, 2003{\natexlab{b}}, \aap, 400, 397

\bibitem[{{Del Zanna} {et~al}\mbox{.}(2007){Del Zanna}, {Zanotti},
  {Bucciantini}, \& {Londrillo}}]{Del-Zanna_Zanotti+07a}
{Del Zanna} L., {Zanotti} O., {Bucciantini} N., {Londrillo} P., 2007, \aap,
  473, 11

\bibitem[{{Duez} {et~al}\mbox{.}(2006){Duez}, {Liu}, {Shapiro}, {Shibata}, \&
  {Stephens}}]{Duez_Liu+06a}
{Duez} M.~D., {Liu} Y.~T., {Shapiro} S.~L., {Shibata} M., {Stephens} B.~C.,
  2006, \prd, 73, 104015

\bibitem[{{Dumbser} \& {Zanotti}(2009)}]{Dumbser_Zanotti09a}
{Dumbser} M., {Zanotti} O., 2009, Journal of Computational Physics, 228, 6991

\bibitem[{{Duncan} \& {Thompson}(1992)}]{Duncan_Thompson92a}
{Duncan} R.~C., {Thompson} C., 1992, \apjl, 392, L9

\bibitem[{{Duttan} \& {Biermann}(2007)}]{Duttan_Biermann07a}
{Duttan} I., {Biermann} P.~L., 2007, {Relativistic Jets in Active Galactic
  Nuclei: Importance of Magnetic Fields}, Aschenbach B. B.~V.~H. G. . L.~B.,
  ed., p. 431

\bibitem[{{Fendt} \& {Memola}(2001)}]{Fendt_Memola01a}
{Fendt} C., {Memola} E., 2001, Astrophysics and Space Science Supplement, 276,
  297

\bibitem[{{Font}, {Stergioulas} \& {Kokkotas}(2000){Font}, {Stergioulas}, \&
  {Kokkotas}}]{Font_Stergioulas+00a}
{Font} J.~A., {Stergioulas} N., {Kokkotas} K.~D., 2000, \mnras, 313, 678

\bibitem[{{Hawley}(2008)}]{Hawley08a}
{Hawley} J., 2008, in APS Meeting Abstracts, p.~H3

\bibitem[{{Kasen} \& {Bildsten}(2010)}]{Kasen_Bildsten10a}
{Kasen} D., {Bildsten} L., 2010, \apj, 717, 245

\bibitem[{{Khanna}(1999)}]{Khanna99a}
{Khanna} R., 1999, in Astronomical Society of the Pacific Conference Series,
  Vol. 178, Stellar Dynamos: Nonlinearity and Chaotic Flows, Ferriz-Mas M.~N.
  .~A., ed., p.~57

\bibitem[{{Khanna} \& {Camenzind}(1992)}]{Khanna_Camenzind92a}
{Khanna} R., {Camenzind} M., 1992, \aap, 263, 401

\bibitem[{{Khanna} \& {Camenzind}(1996{\natexlab{a}})}]{Khanna_Camenzind96a}
---, 1996{\natexlab{a}}, Astrophysical Letters and Communications, 34, 53

\bibitem[{{Khanna} \& {Camenzind}(1996{\natexlab{b}})}]{Khanna_Camenzind96b}
---, 1996{\natexlab{b}}, \aap, 307, 665

\bibitem[{{Komatsu}, {Eriguchi} \& {Hachisu}(1989){Komatsu}, {Eriguchi}, \&
  {Hachisu}}]{Komatsu_Eriguchi+89a}
{Komatsu} H., {Eriguchi} Y., {Hachisu} I., 1989, \mnras, 239, 153

\bibitem[{{Komissarov}(2007)}]{Komissarov07a}
{Komissarov} S.~S., 2007, \mnras, 382, 995

\bibitem[{{Komissarov} \& {Barkov}(2007)}]{Komissarov_Barkov07a}
{Komissarov} S.~S., {Barkov} M.~V., 2007, \mnras, 382, 1029

\bibitem[{{Konigl} \& {Kartje}(1994)}]{Konigl_Kartje94a}
{Konigl} A., {Kartje} J.~F., 1994, \apj, 434, 446

\bibitem[{{Krause} \& {Raedler}(1980)}]{Krause_Raedler80a}
{Krause} F., {Raedler} K.-H., 1980, {Mean-field magnetohydrodynamics and dynamo
  theory}, Goodman L.~J.~\&~Love R.~N., ed.

\bibitem[{{Kulsrud} {et~al}\mbox{.}(1997){Kulsrud}, {Cowley}, {Gruzinov}, \&
  {Sudan}}]{Kulsrud_Cowley+97a}
{Kulsrud} R., {Cowley} S.~C., {Gruzinov} A.~V., {Sudan} R.~N., 1997, \physrep,
  283, 213

\bibitem[{{Kulsrud}(2005)}]{Kulsrud05a}
{Kulsrud} R.~M., 2005, {Plasma physics for astrophysics}

\bibitem[{{Kulsrud} \& {Zweibel}(2008)}]{Kulsrud_Zweibel08a}
{Kulsrud} R.~M., {Zweibel} E.~G., 2008, Reports on Progress in Physics, 71,
  046901

\bibitem[{{Landi} {et~al}\mbox{.}(2008){Landi}, {Londrillo}, {Velli}, \&
  {Bettarini}}]{Landi-Londrillo+08}
{Landi} S., {Londrillo} P., {Velli} M., {Bettarini} L., 2008, Physics of
  Plasmas, 15, 012302

\bibitem[{{Lee}, {Wijers} \& {Brown}(2000){Lee}, {Wijers}, \&
  {Brown}}]{Lee_Wijers+00a}
{Lee} H.~K., {Wijers} R.~A.~M.~J., {Brown} G.~E., 2000, \physrep, 325, 83

\bibitem[{{Lichnerowicz}(1967)}]{Lichnerowicz67a}
{Lichnerowicz} A., 1967, {Relativistic Hydrodynamics and Magnetohydrodynamics},
  Lichnerowicz A., ed.

\bibitem[{{Londrillo} \& {Del Zanna}(2000)}]{Londrillo_Del-Zanna00a}
{Londrillo} P., {Del Zanna} L., 2000, \apj, 530, 508

\bibitem[{{Londrillo} \& {del Zanna}(2004)}]{Londrillo_Del-Zanna04a}
{Londrillo} P., {del Zanna} L., 2004, Journal of Computational Physics, 195, 17

\bibitem[{{Lyons} {et~al}\mbox{.}(2010){Lyons}, {O'Brien}, {Zhang},
  {Willingale}, {Troja}, \& {Starling}}]{Lyons_OBrien+10a}
{Lyons} N., {O'Brien} P.~T., {Zhang} B., {Willingale} R., {Troja} E.,
  {Starling} R.~L.~C., 2010, \mnras, 402, 705

\bibitem[{{Lyutikov}(2003)}]{Lyutikov03b}
{Lyutikov} M., 2003, \mnras, 346, 540

\bibitem[{{Lyutikov}(2006{\natexlab{a}})}]{Lyutikov06b}
---, 2006{\natexlab{a}}, \mnras, 367, 1594

\bibitem[{{Lyutikov}(2006{\natexlab{b}})}]{Lyutikov06a}
---, 2006{\natexlab{b}}, New Journal of Physics, 8, 119

\bibitem[{{Lyutikov} \& {Blackman}(2001)}]{Lyutikov_Blackman01a}
{Lyutikov} M., {Blackman} E.~G., 2001, \mnras, 321, 177

\bibitem[{{Marklund} \& {Clarkson}(2005)}]{Marklund_Clarkson05a}
{Marklund} M., {Clarkson} C.~A., 2005, \mnras, 358, 892

\bibitem[{{Meszaros} \& {Rees}(1997)}]{Meszaros_Rees97a}
{Meszaros} P., {Rees} M.~J., 1997, \apj, 476, 232

\bibitem[{{Metzger} {et~al}\mbox{.}(2011){Metzger}, {Giannios}, {Thompson},
  {Bucciantini}, \& {Quataert}}]{Metzger_Giannios+11a}
{Metzger} B.~D., {Giannios} D., {Thompson} T.~A., {Bucciantini} N., {Quataert}
  E., 2011, \mnras, 413, 2031

\bibitem[{{Metzger}, {Thompson} \& {Quataert}(2007){Metzger}, {Thompson}, \&
  {Quataert}}]{Metzger_Thompson+07a}
{Metzger} B.~D., {Thompson} T.~A., {Quataert} E., 2007, \apj, 659, 561

\bibitem[{{Miralles}, {Pons} \& {Urpin}(2000){Miralles}, {Pons}, \&
  {Urpin}}]{Miralles_Pons+00a}
{Miralles} J.~A., {Pons} J.~A., {Urpin} V.~A., 2000, \apj, 543, 1001

\bibitem[{{Miralles}, {Pons} \& {Urpin}(2002){Miralles}, {Pons}, \&
  {Urpin}}]{Miralles_Pons+02a}
---, 2002, \apj, 574, 356

\bibitem[{{Mizuno} {et~al}\mbox{.}(2011){Mizuno}, {Pohl}, {Niemiec}, {Zhang},
  {Nishikawa}, \& {Hardee}}]{Mizuno_Pohl+11a}
{Mizuno} Y., {Pohl} M., {Niemiec} J., {Zhang} B., {Nishikawa} K.-I., {Hardee}
  P.~E., 2011, \apj, 726, 62

\bibitem[{{Moffatt}(1978)}]{Moffatt78a}
{Moffatt} H.~K., 1978, {Magnetic field generation in electrically conducting
  fluids}, Moffatt H.~K., ed.

\bibitem[{{Munz} {et~al}\mbox{.}(2000){Munz}, {Omnes}, {Schneider},
  {Sonnendrücker}, \& {Voß}}]{Munz_Omnes+00a}
{Munz} C.-D., {Omnes} P., {Schneider} R., {Sonnendrücker} E., {Voß} U., 2000,
  Journal of Computational Physics, 161, 484

\bibitem[{{Murakami}(1999)}]{Murakami99a}
{Murakami} T., 1999, Astronomische Nachrichten, 320, 257

\bibitem[{{Obergaulinger} {et~al}\mbox{.}(2006){Obergaulinger}, {Aloy},
  {Dimmelmeier}, \& {M\"{u}ller}}]{Obergaulinger_Aloy+06a}
{Obergaulinger} M., {Aloy} M.~A., {Dimmelmeier} H., {M\"{u}ller} E., 2006,
  \aap, 457, 209

\bibitem[{{Palenzuela} {et~al}\mbox{.}(2009){Palenzuela}, {Lehner}, {Reula}, \&
  {Rezzolla}}]{Palenzuela_Lehner+09a}
{Palenzuela} C., {Lehner} L., {Reula} O., {Rezzolla} L., 2009, \mnras, 394,
  1727

\bibitem[{{Pareschi} \& {Russo}(2005)}]{Pareschi_Russo05a}
{Pareschi} L., {Russo} G., 2005, Journal of Scientific Computing, 25, 129

\bibitem[{{Pariev}, {Colgate} \& {Finn}(2007){Pariev}, {Colgate}, \&
  {Finn}}]{Pariev_Colgate+07a}
{Pariev} V.~I., {Colgate} S.~A., {Finn} J.~M., 2007, \apj, 658, 129

\bibitem[{{Parker}(1955)}]{Parker55a}
{Parker} E.~N., 1955, \apj, 122, 293

\bibitem[{{Parker}(1987)}]{Parker87a}
---, 1987, \solphys, 110, 11

\bibitem[{{Parker}(2009)}]{Parker09a}
---, 2009, \ssr, 144, 15

\bibitem[{{Penna} {et~al}\mbox{.}(2010){Penna}, {McKinney}, {Narayan},
  {Tchekhovskoy}, {Shafee}, \& {McClintock}}]{Penna_McKinney+10a}
{Penna} R.~F., {McKinney} J.~C., {Narayan} R., {Tchekhovskoy} A., {Shafee} R.,
  {McClintock} J.~E., 2010, \mnras, 408, 752

\bibitem[{{Rezzolla} {et~al}\mbox{.}(2011){Rezzolla}, {Giacomazzo}, {Baiotti},
  {Granot}, {Kouveliotou}, \& {Aloy}}]{Rezzolla_Giacomazzo+11a}
{Rezzolla} L., {Giacomazzo} B., {Baiotti} L., {Granot} J., {Kouveliotou} C.,
  {Aloy} M.~A., 2011, \apjl, 732, L6+

\bibitem[{{Roberts} \& {Soward}(1992)}]{Roberts_Soward92a}
{Roberts} P.~H., {Soward} A.~M., 1992, Annual Review of Fluid Mechanics, 24,
  459

\bibitem[{{Ruediger} {et~al}\mbox{.}(2008){Ruediger}, {Schultz}, {Mond}, \&
  {Shalybkov}}]{Ruediger_Schultz+08a}
{Ruediger} G., {Schultz} M., {Mond} M., {Shalybkov} D.~A., 2008, ArXiv e-prints

\bibitem[{{Sato}(1999)}]{Sato99a}
{Sato} T., 1999, Journal of Plasma and Fusion Research, 75, 7

\bibitem[{{Sauty}, {Tsinganos} \& {Trussoni}(2002){Sauty}, {Tsinganos}, \&
  {Trussoni}}]{Sauty_Tsinganos+02a}
{Sauty} C., {Tsinganos} K., {Trussoni} E., 2002, in Lecture Notes in Physics,
  Berlin Springer Verlag, Vol. 589, Relativistic Flows in Astrophysics,
  A.~W.~Guthmann M.~Georganopoulos A.~M. . K.~M., ed., p.~41

\bibitem[{{Shukurov}(2002)}]{Shukurov02a}
{Shukurov} A., 2002, \apss, 281, 285

\bibitem[{{Tchekhovskoy}, {Narayan} \& {McKinney}(2011){Tchekhovskoy},
  {Narayan}, \& {McKinney}}]{Tchekhovskoy_Narayan+11a}
{Tchekhovskoy} A., {Narayan} R., {McKinney} J.~C., 2011, \mnras, 418, L79

\bibitem[{{Thompson}(1994)}]{Thompson94a}
{Thompson} C., 1994, \mnras, 270, 480

\bibitem[{{Thompson} \& {Duncan}(1996)}]{Thompson_Duncan96a}
{Thompson} C., {Duncan} R.~C., 1996, \apj, 473, 322

\bibitem[{{Tomimatsu}(2000)}]{Tomimatsu00a}
{Tomimatsu} A., 2000, \apj, 528, 972

\bibitem[{{Uzdensky}(2011)}]{Uzdensky11a}
{Uzdensky} D.~A., 2011, \ssr, 160, 45

\bibitem[{{Uzdensky} \&
  {MacFadyen}(2007{\natexlab{a}})}]{Uzdensky_MacFadyen07a}
{Uzdensky} D.~A., {MacFadyen} A.~I., 2007{\natexlab{a}}, \apj, 669, 546

\bibitem[{{Uzdensky} \&
  {MacFadyen}(2007{\natexlab{b}})}]{Uzdensky_MacFadyen07b}
---, 2007{\natexlab{b}}, Physics of Plasmas, 14, 056506

\bibitem[{{van Putten} \& {Levinson}(2003)}]{van-Putten_Levinson03a}
{van Putten} M.~H.~P.~M., {Levinson} A., 2003, \apj, 584, 937

\bibitem[{{Vlahakis} \& {K\"{o}nigl}(2001)}]{Vlahakis_Konigl01a}
{Vlahakis} N., {K\"{o}nigl} A., 2001, \apjl, 563, L129

\bibitem[{{Woosley}(2010)}]{Woosley10a}
{Woosley} S.~E., 2010, \apjl, 719, L204

\bibitem[{{Yu}(2011)}]{Yu11a}
{Yu} C., 2011, \apj, 738, 75

\bibitem[{{Zanotti} \& {Dumbser}(2011)}]{Zanotti_Dumbser11a}
{Zanotti} O., {Dumbser} M., 2011, \mnras, 418, 1004

\bibitem[{{Zeldovich} \& {Ruzma\v{i}kin}(1987)}]{Zeldovich_Ruzmaikin87a}
{Zeldovich} Y.~B., {Ruzma\v{i}kin} A.~A., 1987, Soviet Physics Uspekhi, 30, 494

\bibitem[{{Zhang}, {MacFadyen} \& {Wang}(2009){Zhang}, {MacFadyen}, \&
  {Wang}}]{Zhang_MacFadyen+09a}
{Zhang} W., {MacFadyen} A., {Wang} P., 2009, \apjl, 692, L40

\end{thebibliography}
\bibliographystyle{mn2e}

\label{lastpage}

\end{document}